\documentclass[10pt,draftclsnofoot,journal,letterpaper,onecolumn]{IEEEtran}

\usepackage{cite}
\usepackage{graphicx}
\usepackage{subfigure}
\graphicspath{{}}

\title{Subwavelength Nanopatch Cavities for Semiconductor Plasmon Lasers} 
\author{ Christina Manolatou, Farhan Rana \\
School of Electrical and Computer Engineering, \\ 
Cornell University, Ithaca, NY 14853}

\begin{document}

\maketitle

\begin{abstract}
We propose and analyze a family of nanoscale cavities for electrically-pumped surface-emitting semiconductor lasers that use surface plasmons to provide optical mode confinement in cavities which have dimensions in the 100-300 nm range. The proposed laser cavities are in many ways nanoscale optical versions of micropatch antennas that are commonly used at microwave/RF frequencies. Surface plasmons are not only used for mode confinement but also for output beam shaping to realize single-lobe far-field radiation patterns with narrow beam waists from subwavelength size cavities. We identify the cavity modes with the largest quality factors and modal gain, and show that in the near-IR wavelength range (1.0-1.6 $\mu$m) cavity losses (including surface plasmon losses) can be compensated by the strong mode confinement in the gain region provided by the surface plasmons themselves and the required material threshold gain values can be smaller than 700 cm$^{-1}$. 
\end{abstract}


\IEEEpeerreviewmaketitle

\section{Introduction}
Electrically pumped semiconductor lasers with nanometer scale optical cavities could be important for applications that benefit from ultrasmall coherent light sources, such as on-chip optical interconnects, dense photonic VLSI circuits, and biological or chemical sensors for micro- and nano-systems. Two questions that are interesting in this context are: (a) what are the smallest achievable dimensions of an electrically pumped semiconductor laser consistent with the current material and fabrication constrains, and (b) what is the quality of output beams shapes obtainable from subwavelength laser cavities. 

In the past few years much progress has been made in tightly confining light in high quality factor optical microcavities and in defects in 1D and 2D photonic crystals~\cite{vahala,joan,yariv}. Modal volumes close to the diffraction limit of $(\lambda/2n)^{3}$ (where $\lambda$ is the mode wavelength and $n$ is the refractive index seen by the mode) have been achieved in some of these structures~\cite{vahala,joan,yariv}. Optically and electrically pumped photonic crystal defect lasers with modal volumes few times the diffraction limit have also been demonstrated~\cite{noda,painter,bhat}. Feedback structures, such as Bragg reflectors and photonic crystals, needed to achieve such small modal volumes made the overall size of the laser structure several times larger than the wavelength since at least a few periods of the feedback structure were required for adequate mode confinement~\cite{vahala,joan,yariv,noda,painter,bhat}. Plasmonic structures for photonic applications have been extensively studied in the last few years~\cite{coyle,dereux,ozbay,orein,hung,fainman,larsen,atwater1,atwater2,maier1,maier2}. The large wavevector values of plasmon-polaritons near the surface plasmon resonance frequency have been used to achieve subwavelength device dimensions. Surface plasmon confined optical modes in waveguides have exhibited modal loss values ranging from 0.3 dB/$\mu$m to 30 dB/$\mu$m in the visible-to-near-IR wavelength range~\cite{larsen,atwater1}. While large wavevector values, and small device sizes (compared to the free-space wavelength), are possible for frequencies near the surface plasmon resonance frequency (which corresponds to wavelengths in the 0.4-0.6 $\mu$m range for most important metals, such as Silver and Gold), the losses are also higher at frequencies close to the surface plasmon resonance frequency~\cite{dereux,ozbay}. In the near-IR 1.0-1.6 $\mu$m wavelength range, although the losses are smaller, the wavevector values of plasmon-polaritons are also smaller~\cite{dereux,ozbay}. Plasmon propagation in waveguides coupled to gain media has also been studied theoretically and values between 500 cm$^{-1}$ and 5000 cm$^{-1}$ for the material gain required for lossless propagation have been reported~\cite{fainman,alam}. An advantage of operating at frequencies much smaller than the surface plasmon resonance frequency is that the plasmon fields are not strongly confined near the surface of the metal and can therefore have significant overlap with an external gain medium. For realizing lasing in surface plasmon confined nanoscale optical cavities, the gain needs to not only compensate for intrinsic cavity losses (including surface plasmon losses) but also for losses due to external radiation. In addition, the output radiation patterns need to be well behaved for practical applications. In the mid-IR and far-IR wavelength range, where surface plasmon losses are considerably smaller compared to those at visible and near-IR wavelengths, surface plasmon mode confinement in dual-metal waveguides has been used to achieve lasing in quantum cascade devices~\cite{capasso, williams1, capasso2}. However, these laser structures did not have subwavelength sizes in all three dimensions.

In this paper, we propose and analyze a family of nanoscale optical cavities for electrically-pumped surface-emitting near-IR semiconductor lasers - semiconductor nanopatch lasers (or SNLs) - that have dimensions in the few hundred nanometer range and cavity volumes (not just modal volumes) approaching $(\lambda/2n)^{3}$~\cite{rana}. Surface plasmons are used not only for mode confinement but also for output beam shaping to realize single-lobe far-field output radiation patterns with narrow beam waists. The lasers discussed here are in many ways nanoscale optical versions of micropatch antennas (or microstrip patch antennas) that are commonly used at microwave/RF frequencies~\cite{balannis}. We show that in the near-IR wavelength region (1.0-1.6 $\mu$m) cavity losses in nanopatch lasers (including surface plasmon losses) can be compensated by gain from conventional III-V materials because surface plasmons can themselves be used to provide large overlap of the cavity mode with the gain region. The material threshold gain values needed to achieve lasing can be smaller than 700 cm$^{-1}$. Despite surface plasmon losses, the external radiation efficiencies are between 10$\%$-30$\%$. Compared to all-dielectric microcavity semiconductor lasers reported in the literature~\cite{noda,painter,bhat,fujita}, the proposed nanopatch lasers have much smaller cavity volumes and well behaved output beam shapes. 

In Section~\ref{secstr} we discuss the basic structure of nanopatch laser cavities. Section~\ref{secoptical} describes the general characteristics of the modes most favorable for achieving lasing in subwavelength nanopatch cavities. Section~\ref{secsim} is the main section of this paper and presents simulation results for cavity modes and corresponding radiation fields. The focus is mainly on circular nanopatch cavities as they are the easiest to analyze computationally and the main characteristics of other nanopatch laser cavities are similar. Simulation results for the threshold material gain and threshold current values required for achieving lasing are presented in Section~\ref{secgain}. Finally, the challenges related to the fabrication of nanopatch laser cavities are discussed in Section~\ref{secfinal}.

\section{Nanopatch Laser Structures} \label{secstr}
\begin{figure}[tbp]
\centering
\subfigure{
\includegraphics[width=2.0in]{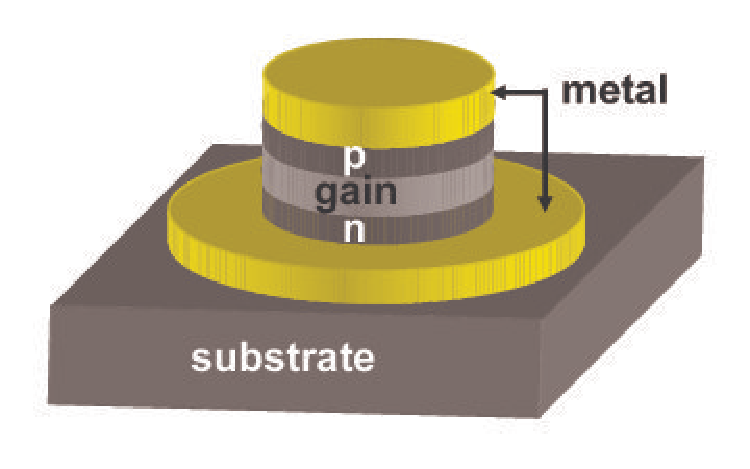}}
\hspace{0.0in}
\subfigure{
\includegraphics[width=2.0in]{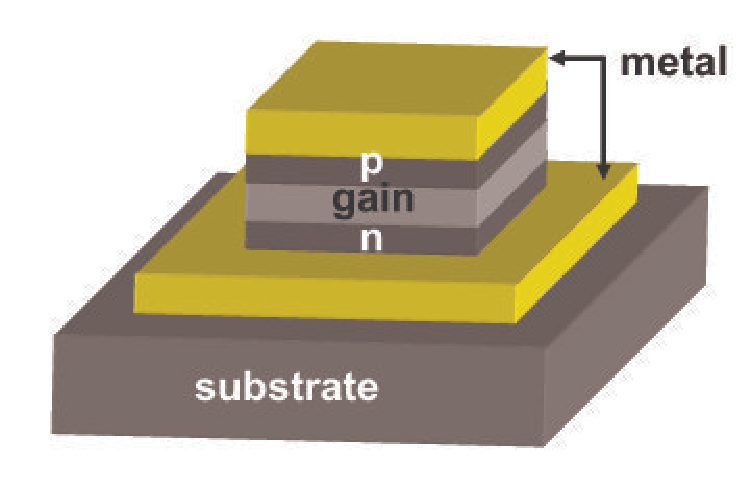}}
\hspace{0.0in}
\subfigure{
\includegraphics[width=2.0in]{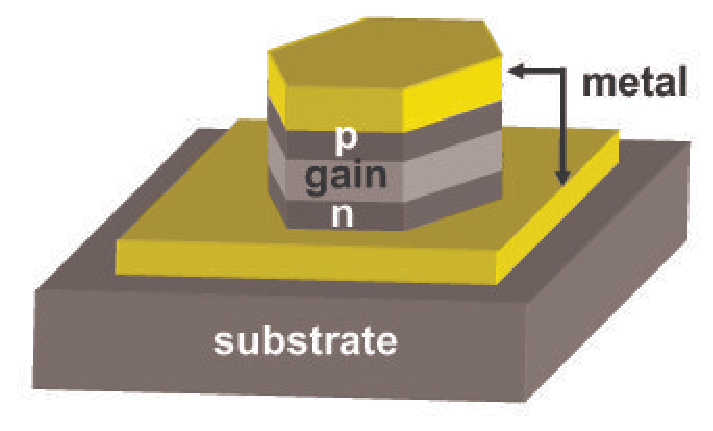}}
\caption{Examples of nanopatch laser structures: (LEFT) a circular nanopatch laser, (MIDDLE) a rectangular nanopatch laser, and (RIGHT) a hexagonal nanopatch laser. The structures are not drawn to scale.}  
\label{nanostr}
\end{figure}

Few examples of nanopatch laser structures are shown in Fig.~\ref{nanostr}. The basic nanopatch laser structure consists of a bulk semiconductor gain medium in the form of a p-i-n heterostructure sandwiched on both sides by metal layers that confine the lasing optical mode via surface plasmons and also serve as electrical contacts for current injection. The light is radiated out from the sides of the cavity. The bottom metal layer acts like an antenna reflector and directs the light radiated out from the sides of the cavity in the upward direction thereby contributing to the surface emission characteristics of nanopatch lasers. Ground planes in micropatch antennas used at microwave/RF frequencies perform similar functions~\cite{balannis}. The radii of the nanopatch lasers are between 100 nm and 300 nm and the heights of the dielectric part of the cavity are between 100 nm and 250 nm. The metal layers are assumed to be thick enough to prohibit light transmission. In the next Section, we discuss the optical modes most suitable for achieving lasing in nanopatch cavities.

\section{Optical Modes in Nanopatch Cavities: Discussion}     \label{secoptical}
The focus in this paper will be only on those modes supported by nanopatch cavities that exhibit the smallest surface plasmon and radiation losses for subwavelength cavity dimensions, require the smallest cavity dimensions for a given lasing frequency, and have the largest overlaps with the gain region. Numerical simulations (to be discussed below) show that in all the nanopatch cavities discussed in this paper the most favorable modes with respect to the above figures of merit are similar in shape and profile. Below, we discuss these modes in more detail in the context of circular nanopatch lasers. 

\begin{figure}[t]
%\centering
%\includegraphics[width=2.5in,angle=-90]{tmmodes}    
\subfigure{
\includegraphics[width=2.0in]{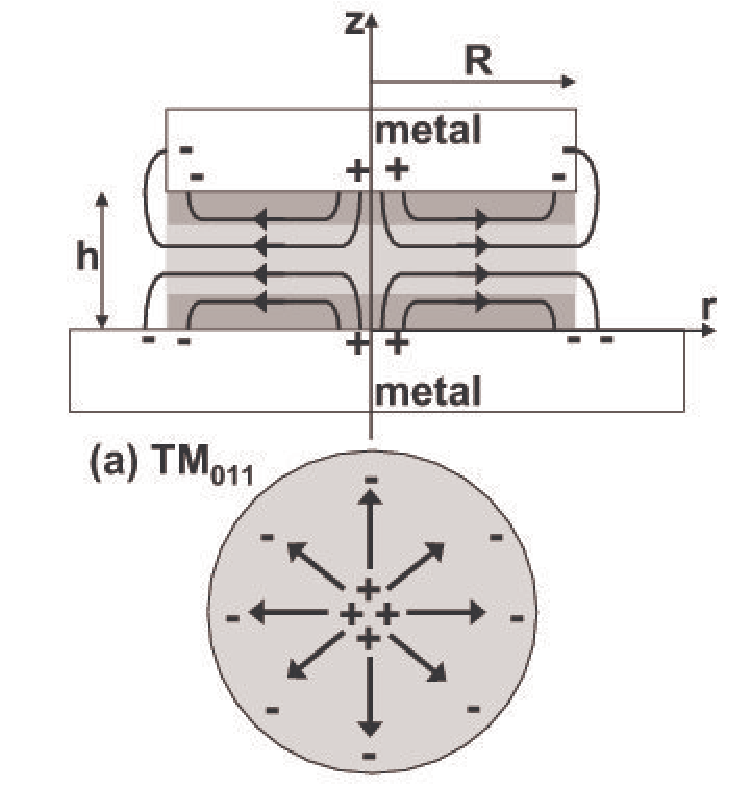}}
\hspace{0.0in}
\subfigure{
\includegraphics[width=2.0in]{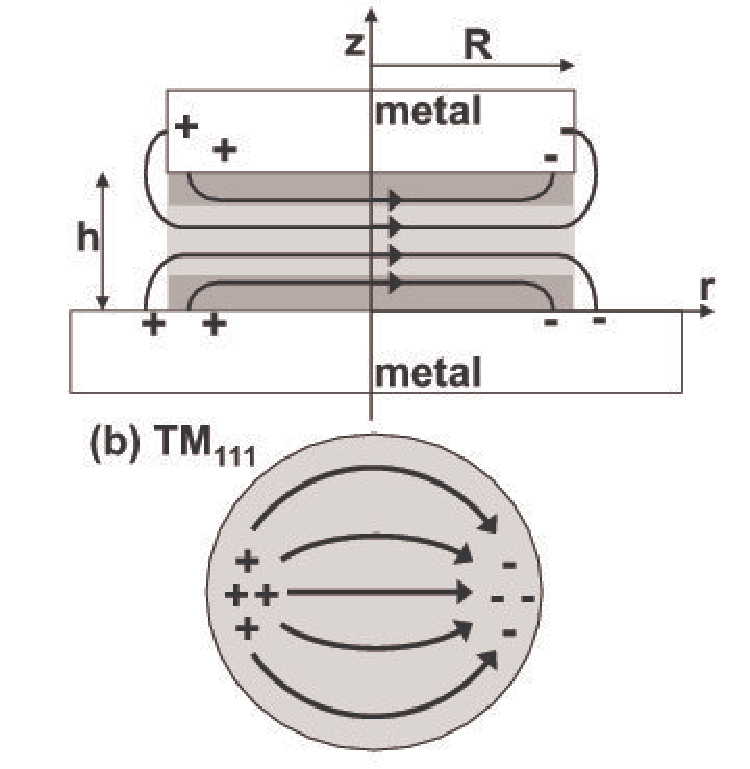}}
\hspace{0.0in}
\subfigure{
\includegraphics[width=2.0in]{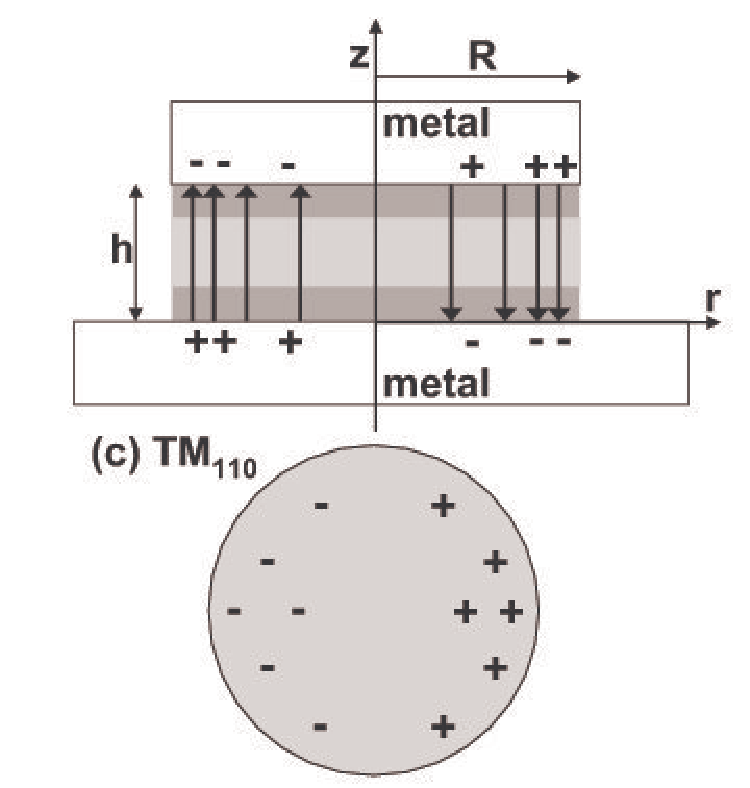}}
\hspace{0.0in}
\caption{Fields lines for (a) TM$_{011}$, (b) TM$_{111}$, and (c) TM$_{110}$ modes of a circular nanopatch cavity. The figures on the top show the side view for the fields lines and the figures on the bottom show the top view in the $z=h/2$ plane.} 
\label{tmmodes}
\end{figure}

\subsection{Optical Modes in Circular Nanopatch Cavities}
Circular nanopatch cavities are the easiest to analyze both computationally and analytically. The continuous angular symmetry implies that the angular dependence of the modal field amplitude can be written as $\exp(\pm i m \phi)$ where $m$ is an integer. For $m \ne 0$ there are two degenerate modes for each value of $m$ corresponding to positive and negative values of the sign in the exponential. Equivalently, the angular dependence of the modes can be assumed to be $\cos(m\phi)$ or $\sin(m\phi)$ instead of $\exp(\pm i m \phi)$. With reference to the vertical direction, the modes can be further classified as quasi-TM or quasi-TE~\cite{balannis}. The TE modes do not couple well with the surface plasmons and require larger cavity dimensions for the same mode wavelength compared to the TM modes. Therefore, TE modes will not be considered in this paper. The TM modes of circular nanopatch cavities bear a close resemblance to the TM$_{mnp}$ modes of circular micropatch antennas which inside the cavity can be written as~\cite{balannis},
\begin{equation} 
E_{z}(r,\phi,z) = E_{o} \, J_{m}(k_{r}r) \, \cos\left( \frac{p\pi}{h} z \right) \, \exp(\pm i m \phi) \label{eqmode1}
\end{equation} 
\begin{equation}
E_{r}(r,\phi,z) = - \frac{p\pi/h}{k_{r}} E_{o} \,  J'_{m}(k_{r}r) \, \sin\left( \frac{p\pi}{h} z \right) \, \exp(\pm i m \phi) \label{eqmode2}
\end{equation}  
\begin{equation}
E_{\phi}(r,\phi,z) = \mp i m \, \frac{p\pi/h}{k_{r}} E_{o} \,  \frac{J_{m}(k_{r}r)}{k_{r}r} \, \sin\left( \frac{p\pi}{h} z \right) \, \exp(\pm i m \phi) \label{eqmode3}
\end{equation}  
where, $h$ is the height of the cavity, $k_{r}^{2} = (\omega_{mnp}/c)^{2} \epsilon - (p \pi/h)^{2}$, $\omega_{mnp}$ is the cavity resonance frequency, and $\epsilon$ is the dielectric constant of the medium inside the cavity. Although the above expressions are not exact for the TM modes of circular nanopatch cavities, they capture the approximate profiles and symmetries of the actual nanopatch cavity modes at near-IR wavelengths that are not close to the surface plasmon frequency. The TM$_{110}$ mode is the mode of choice in circular micropatch antennas at RF/microwave frequencies. This mode has a constant electric field component in the vertical direction and a very small radiation-$Q$. The small radiation-$Q$ makes it suitable for antenna applications but not for achieving lasing. The electric field lines for the TM$_{011}$ and TM$_{111}$ modes of a nanopatch cavity are also depicted in Fig.~\ref{tmmodes}. For these modes, the horizontal component of the electric field is symmetric with respect to a horizontal plane passing through the middle of the cavity and the vertical component of the electric field is anti-symmetric with respect to the same plane. These properties of the modes facilitate large mode overlap with the gain region and also result in reduced radiation losses. The reduced radiation losses are due to the partial cancellation of the outgoing radiation as a result of the up-down anti-symmetry of the vertical component of the electric field at the edges of the cavity. For a given frequency, the TM$_{111}$ mode results in smaller cavity dimensions and larger overlap of the mode with the gain region. Higher order TM modes, and in particular TM modes with $m>1$ (whispering gallery modes), do exhibit reduced radiation losses but also require larger cavity dimensions and will therefore not be considered in this work. TM$_{111}$ mode is therefore the most suitable mode for achieving lasing in subwavelength circular nanopatch cavities. As discussed later, the modes most suitable for achieving lasing in other nanopatch cavities with subwavelength dimensions resemble the TM$_{111}$ mode of a circular nanopatch cavity. In the following Section, we present simulation results for the TM$_{111}$-like cavity modes of nanopatch lasers.  

\section{Optical Modes in Nanopatch Cavities: Simulation Results} \label{secsim}

\begin{figure}[t]
\centering
\subfigure{
\includegraphics[width=2.8in]{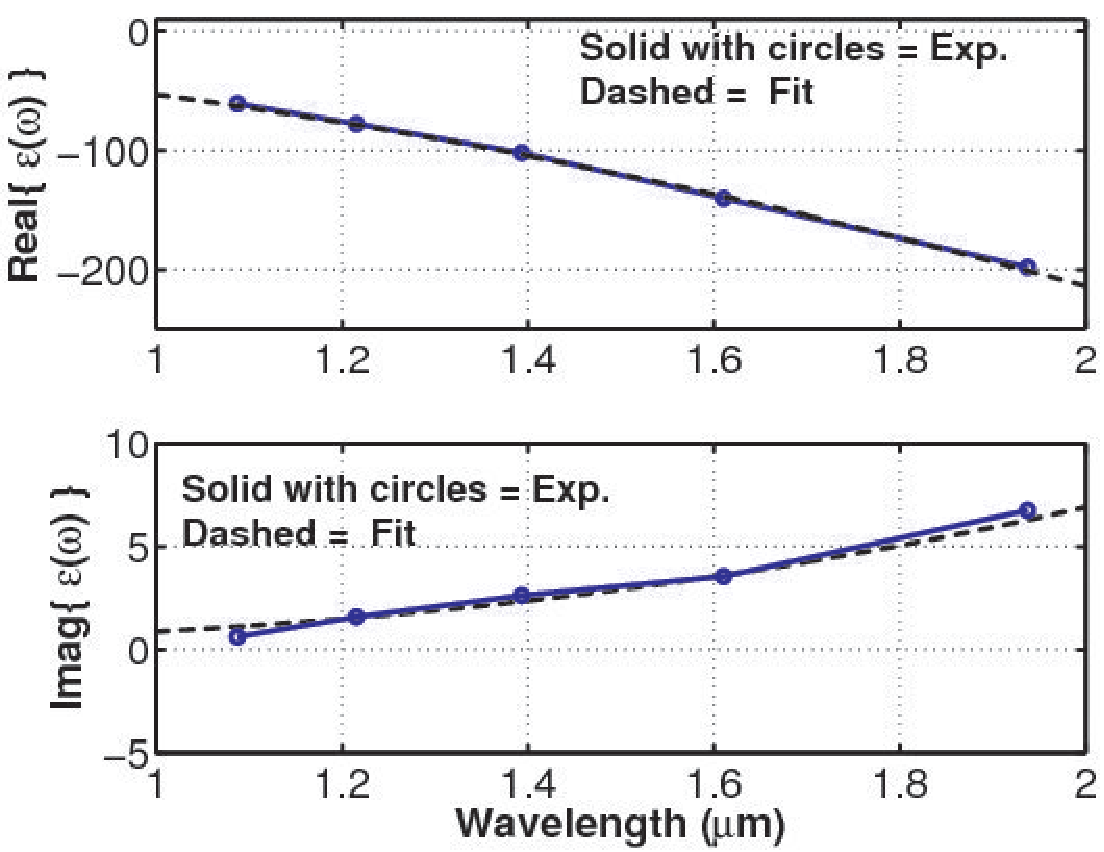}}
\hspace{0.0in}
\subfigure{
\includegraphics[width=3.0in]{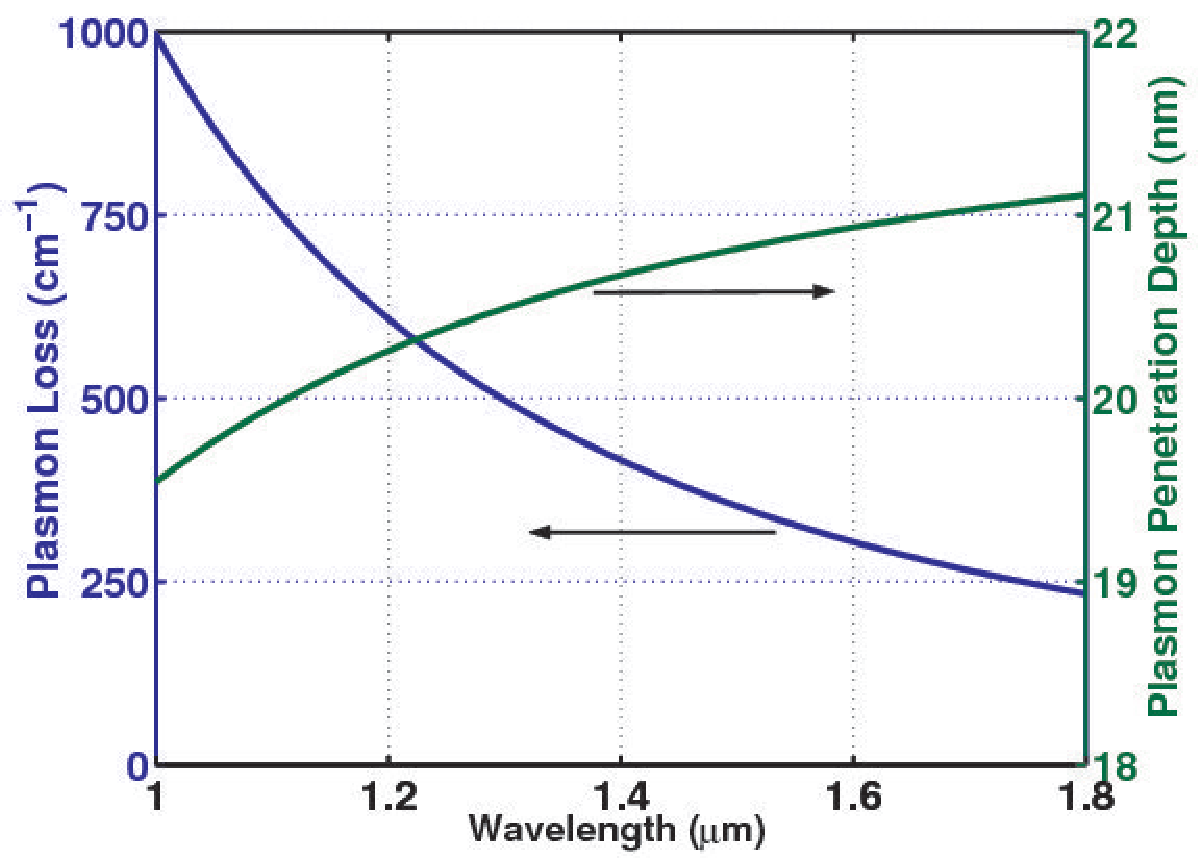}}
\hspace{0.0in}
\caption{(LEFT) The experimental data (SOLID WITH CIRCLES) of Ref.~\cite{jandc} for the real and imaginary parts of the dielectric constant of Silver and the fit (DASHED) using the model plasma dispersion in the 1.0-1.8 $\mu$m wavelength range are shown. (RIGHT) The propagation loss of free plasmons on Silver, and the plasmon field penetration depth in the metal, are shown as a function of the wavelength. The metal is assumed to be surrounded by a material of refractive index 3.3 corresponding to the contact/cladding material (InGaAsP) in nanopatch lasers.} 
\label{epfit}
\end{figure}

\subsection{Simulation Methods and Techniques} \label{sectech}
For numerical simulations a 3D full-vector complex eigenmode solver, with reflectionless perfectly matched layer (PML) boundary conditions for outgoing radiation, is used. A 3D full-vector finite difference time domain (FDTD) simulator with PML boundary conditions is also used~\cite{taflove}. A non-uniform computational mesh is employed and the minimum separation between adjacent mesh points in the metal is 4 nm, which is much smaller than the expected penetration depth of the plasmon field in the metal layers (see Fig.~\ref{epfit}). The refractive index of the semiconductor gain region is assumed to be 3.5 and corresponds to the (approximately wavelength independent) refractive index of InGaAsP gain region in the 1.1-1.6 $\mu$m wavelength range~\cite{adachi}. The refractive index of the contact/cladding layers is assumed to be 3.3 and also wavelength independent (in the 1.0-1.6 $\mu$m wavelength range a refractive index value of 3.3 can always be obtained for an InGaAsP layer by adjusting the composition~\cite{adachi}). The material loss due to free carriers is assumed to be 30 cm$^{-1}$ per $10^{18}$ 1/cm$^{3}$ carrier density for holes and 5 cm$^{-1}$ per $10^{18}$ 1/cm$^{3}$ carrier density for electrons~\cite{coldren}. The n- and p-doped contact/cladding layers on each side of the gain region are both assumed to have carrier densities of $4\times 10^{18}$ 1/cm$^{3}$. 

In all the simulations, the metal is assumed to be Silver and the experimentally measured wavelength dependent plasma dispersion of Silver reported by Johnson and Christy~\cite{jandc} is incorporated in the simulations. In the FDTD method, the plasma dispersion is included by fitting the experimentally measured dispersion with the model plasma dispersion relation,
\begin{equation}
\epsilon(\omega) = \epsilon_{\infty} - \frac{\omega_{p}^{2}}{\omega \, (\omega + i/\tau)}
\end{equation}    
and then adding to the FDTD equations an additional equation for the material polarization current density according to the model plasma dispersion, as described in Ref.~\cite{taflove}. The values of $\epsilon_{\infty}$, $\omega_{p}$, and $\tau$ that best fit the experimental data in the 1.0-1.8 $\mu$m wavelength range are $1.0$, $1.38\times 10^{16}$ rad/s, and 33 fs, respectively. The quality of the fit is shown in Fig.~\ref{epfit}. Fig.~\ref{epfit} also shows the propagation loss of free plasmons on Silver, and the plasmon field penetration depth in the metal, as a function of wavelength and obtained using the model plasmon dispersion relation. These results will be used in the discussion that follows. In the case of the eigenmode solver, the eigenequation that needs to be solved numerically is,
\begin{equation}
 \nabla \times \frac{1}{\epsilon(\vec{r},\omega)} \, \nabla \times \vec{H}(\vec{r}) = \frac{\omega^{2}}{c^{2}} \, \vec{H}(\vec{r})
\end{equation}   
Since the dielectric constant also depends on the mode frequency, an iterative scheme is implemented in which the value of the dielectric constant is updated in every iteration until convergence is obtained. With PML boundary conditions, the eigenmode solver gives a complex mode frequency where the imaginary part is related to the photon lifetime in the cavity. In the FDTD method, the photon lifetime is obtained directly by observing the decay of the field inside the cavity with time. The photon lifetime $\tau_{p}$ is related to the cavity quality factor $Q$ by, $Q=\omega \tau_{p}$, where $\omega$ is the mode frequency. The photon lifetime $\tau_{p}$ can be written as,
\begin{equation}
\frac{1}{\tau_{p}} = \frac{1}{\tau_{i}} + \frac{1}{\tau_{r}} \label{eqphtimes}
\end{equation}
where, $\tau_{i}$ and  $\tau_{r}$ represent the intrinsic cavity loss and the radiation loss, respectively. The value of $\tau_{r}$ is found by calculating the total power $P_{r}$ radiated by the cavity (in the far-field) and dividing it by the total energy $W_{e}$ stored in the cavity, 
\begin{eqnarray}
\frac{1}{\tau_{r}} & = & \frac{P_{r}}{W_{e}} = \frac{ 2 {\rm Real} \left[ \int \int_{surface}\, \left( \vec{E}(\vec{r})\times \vec{H^{*}}(\vec{r}) \right).d\vec{s} \right] }{ {\rm Real} \left[ \int \int \int_{cavity} \, \epsilon_{o} \, (\partial\omega \epsilon/\partial \omega)\, \vec{E}(\vec{r}). \vec{E^{*}}(\vec{r}) + \mu_{o} \vec{H}(\vec{r}). \vec{H^{*}}(\vec{r}) \,\, dV \right]} \nonumber \\
& = & \frac{ {\rm Real} \left[ \int \int_{surface}\, \left( \vec{E}(\vec{r})\times \vec{H^{*}}(\vec{r}) \right).d\vec{s} \right] }{ {\rm Real} \left[ \int \int \int_{cavity} \, \epsilon_{o} \, (\epsilon + (\omega/2)\, \partial \epsilon/\partial \omega)\, \vec{E}(\vec{r}). \vec{E^{*}}(\vec{r}) \,\, dV \right]}
\end{eqnarray}
It needs to be pointed out here that a slightly different expression for the electromagnetic energy density in a dispersive and lossy material is given in Ref.~\cite{maier1}. However, for the parameter values chosen for the dielectric dispersion the difference is negligible. The integral for the total power is performed over a hemispherical surface of radius sufficiently large that the result is not affected by the near-fields of the cavity. The volume integral for the energy stored in the cavity is performed over a region slightly larger than just the dielectric and the metal regions of the cavity to include the fringing fields in the air region as well. We have found that in all the cases considered in this paper more than 90$\%$ of the field energy is stored in the dielectric and the metal regions of the cavity. Once $\tau_{r}$ is found, $\tau_{i}$ is obtained using Equation~(\ref{eqphtimes}). The external radiation efficiency $\eta_{ext}$ of the laser is defined as,
\begin{equation}
 \eta_{ext} = \frac{\rm Radiation \, \, Loss}{\rm Total \,\, Loss} = \frac{\tau_{i}}{\tau_{i} + \tau_{r}}
\end{equation} 
Finally, the fraction $\Gamma_{e}$ of the cavity energy in the gain region is calculated as,
\begin{equation}
\Gamma_{e} = \frac{ {\rm Real} \left[ \int \int \int_{gain} \, \epsilon_{o} \, (\epsilon + (\omega/2) \partial \epsilon/\partial \omega)\, \vec{E}(\vec{r}). \vec{E^{*}}(\vec{r}) \,\, dV \right] }{ {\rm Real} \left[ \int \int \int_{cavity} \, \epsilon_{o} \, (\epsilon + (\omega/2) \partial \epsilon/\partial \omega)\, \vec{E}(\vec{r}). \vec{E^{*}}(\vec{r}) \,\, dV \right]}
\end{equation}
The results obtained from the FDTD and the eigensolution techniques were compared for several different cavities and the maximum difference in the calculated mode frequencies and photon lifetimes was found to be less than 2$\%$. The numerical error in the simulation results is expected to be of the same order. 

\begin{figure}[tbp]
\centering
\subfigure{
\includegraphics[width=1.95in,angle=-0]{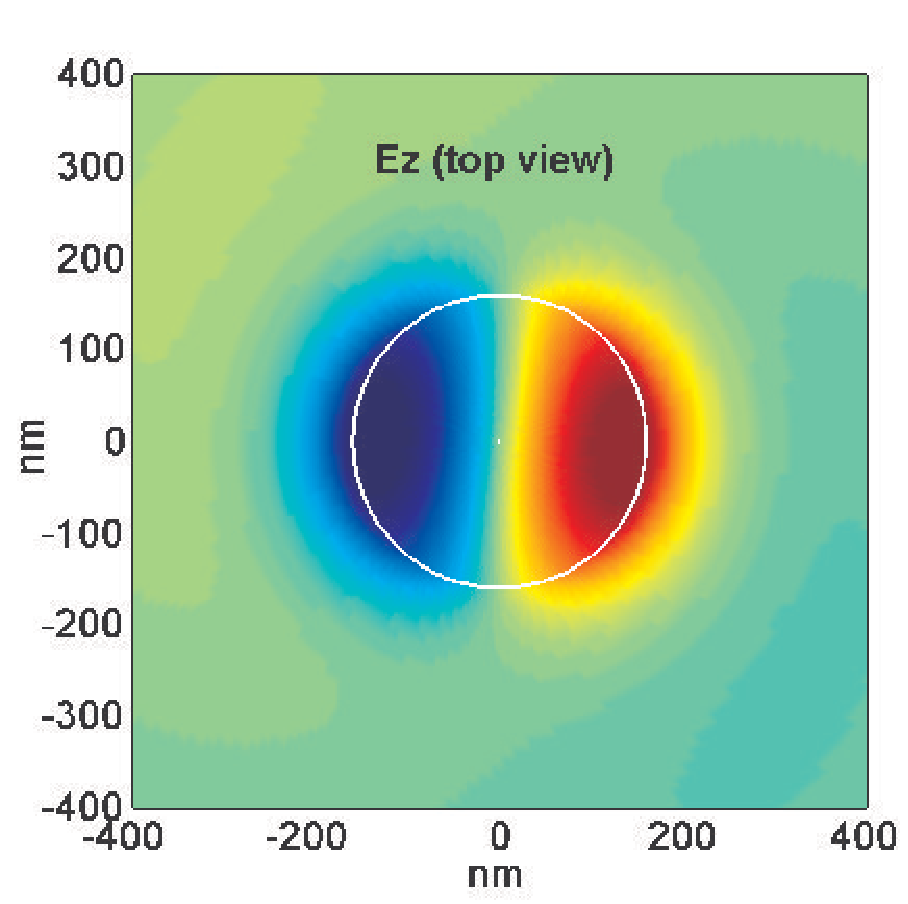}}    
\hspace{0.0in}
\subfigure{
\includegraphics[width=2.0in,angle=-0]{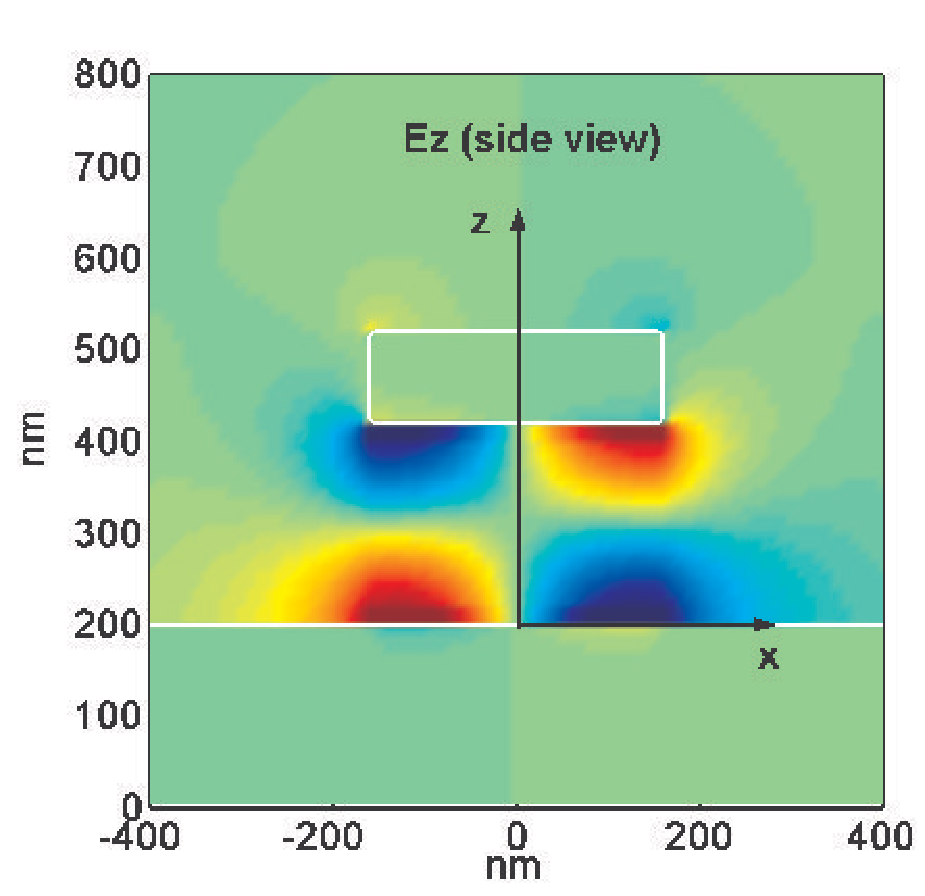}}    
\hspace{0.0in}
\subfigure{
\includegraphics[width=2.0in,angle=-0]{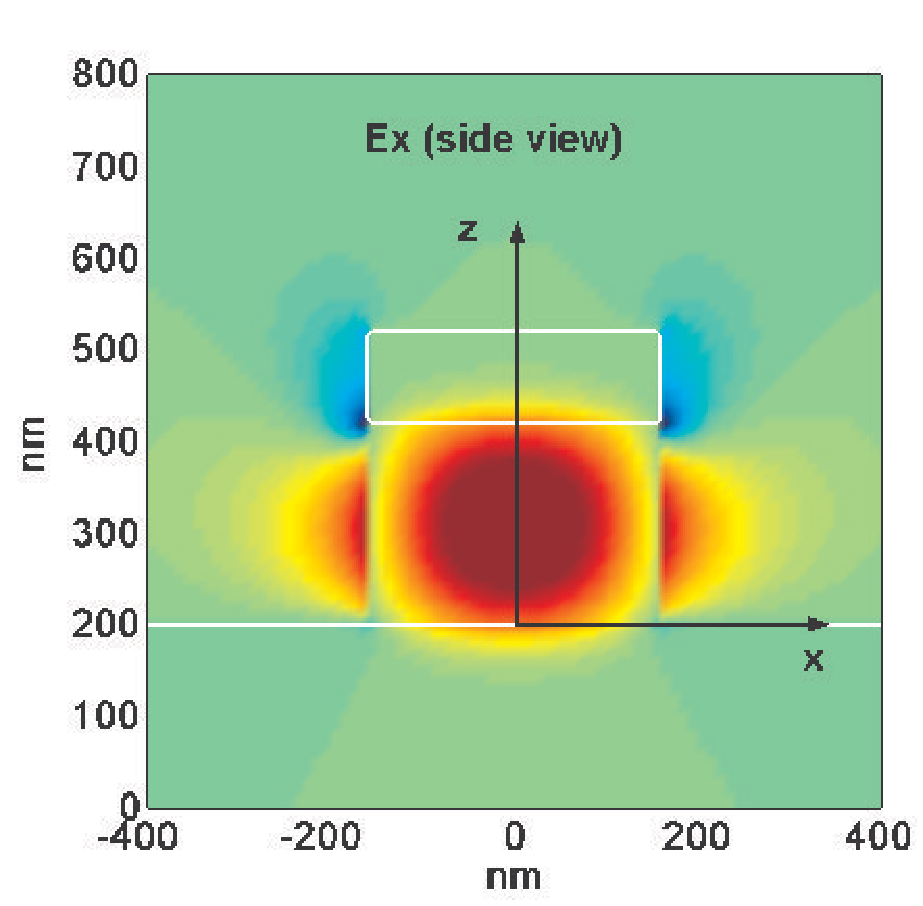}}
\caption{The computed $z$-component(top view) (LEFT), $z$-component(side view) (MIDDLE), and  the $x$-component(side view) of the electric field for the TM$_{111}$ mode of a circular nanopatch cavity of height 220 nm and radius 160 nm are shown. The $z$-component (top view) is shown in a horizontal plane just below the top metal. The $z$- and the $x$-components (side view) are shown in a vertical plane passing through the center of the cavity. The contrast has been reduced to show the weak fringing fields in the air region.}
\label{figcirsnl}
\end{figure}

\begin{figure}[tbp]
\centering
\subfigure{
\includegraphics[width=3.0in]{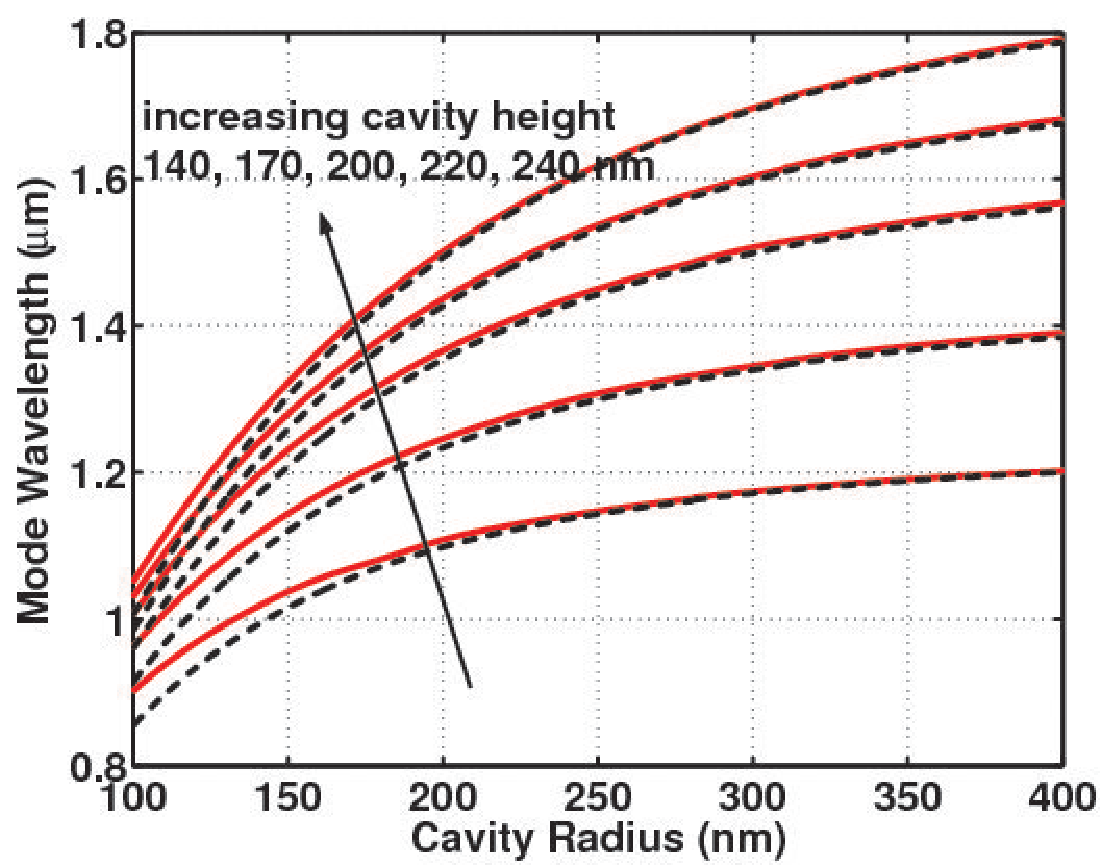}}
\hspace{0.1in}
\subfigure{
\includegraphics[width=3.0in]{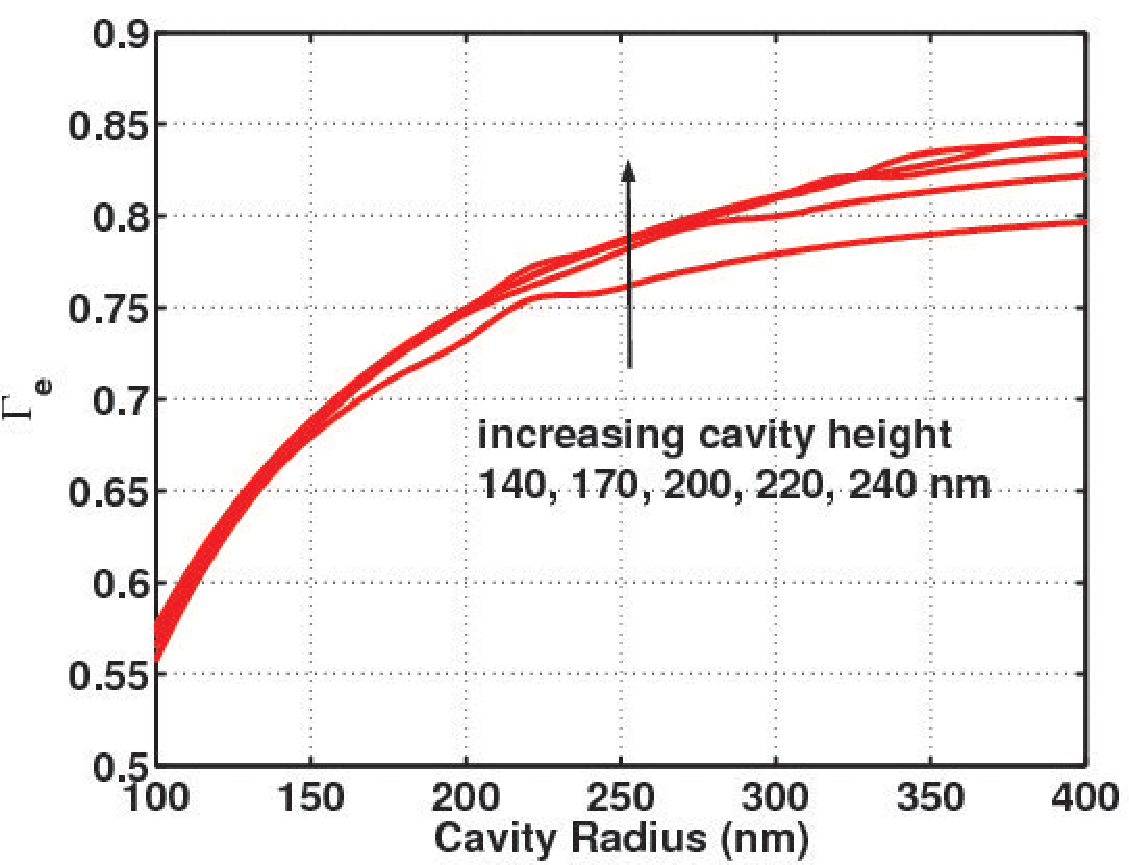}}
\caption{(LEFT) The computed wavelengths for the TM$_{111}$ mode of a circular nanopatch cavity are plotted as a function of the cavity radius for different cavity heights. The SOLID lines are the results from numerical simulations and the DASHED lines correspond to the analytical expression in Eq.(\ref{eqformula}). (RIGHT) The fraction $\Gamma_{e}$ of the cavity electromagnetic energy in the gain region for the TM$_{111}$ mode is plotted as a function of the cavity radius for different cavity heights. The cavity heights are 140, 170, 200, 220, and 240 nm in each case.}  
\label{figradwavgamma}
\end{figure}
\begin{figure}[tbp]
\centering
\subfigure{
\includegraphics[width=3.0in]{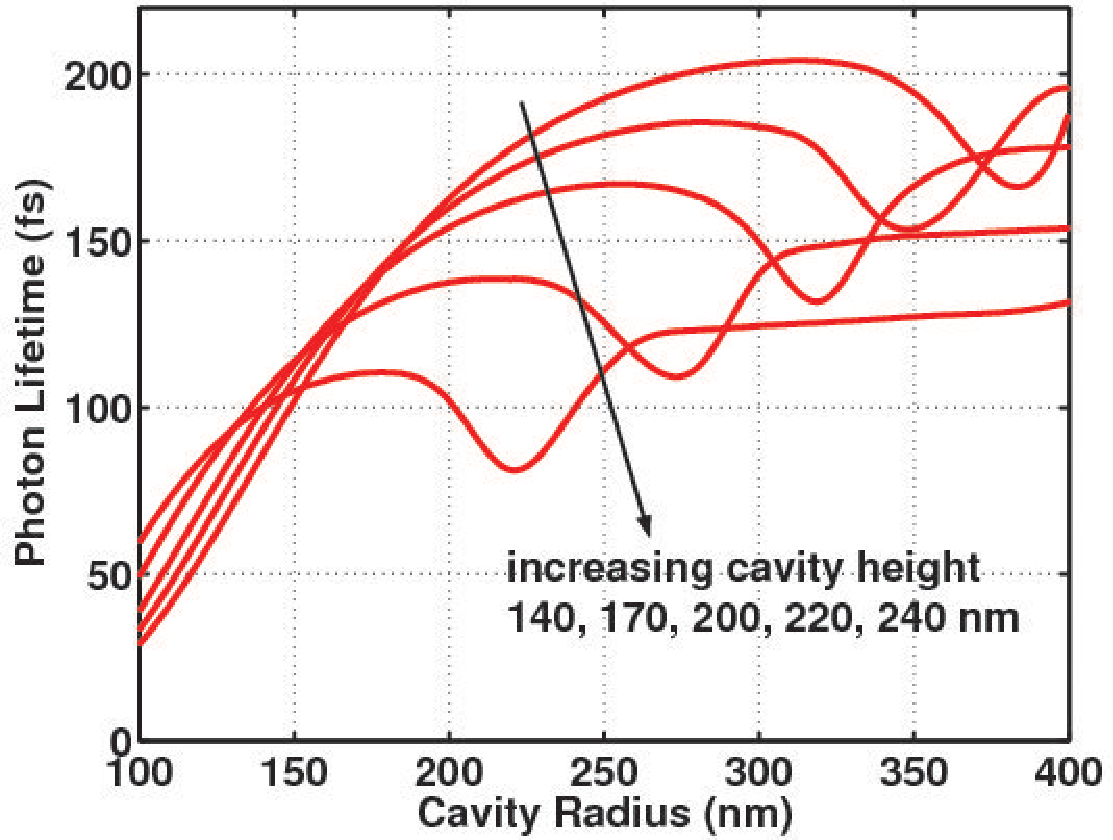}}
\hspace{0.1in}
\subfigure{
\includegraphics[width=3.0in]{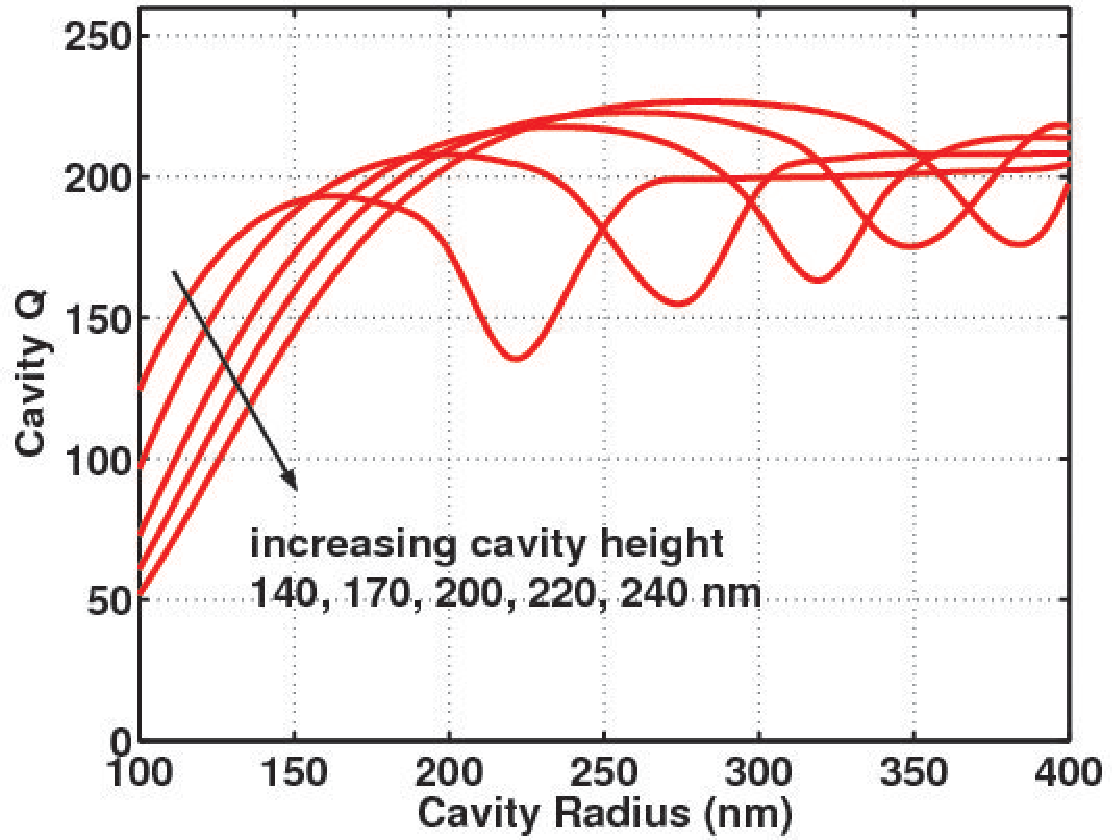}}
\caption{The computed cavity photon lifetimes (LEFT), and the fraction $\Gamma_{e}$ of the cavity electromagnetic energy in the gain region (RIGHT), for TM$_{111}$ modes of circular nanopatch cavities are plotted as a function of the cavity radius for different cavity heights. The cavity heights are 140, 170, 200, 220, and 240 nm. The corresponding mode wavelengths are given in Fig.~\ref{figradwavgamma}. }  
\label{figwavtpq}
\end{figure}
\begin{figure}[tbp]
\centering
\subfigure{
\includegraphics[width=3.0in]{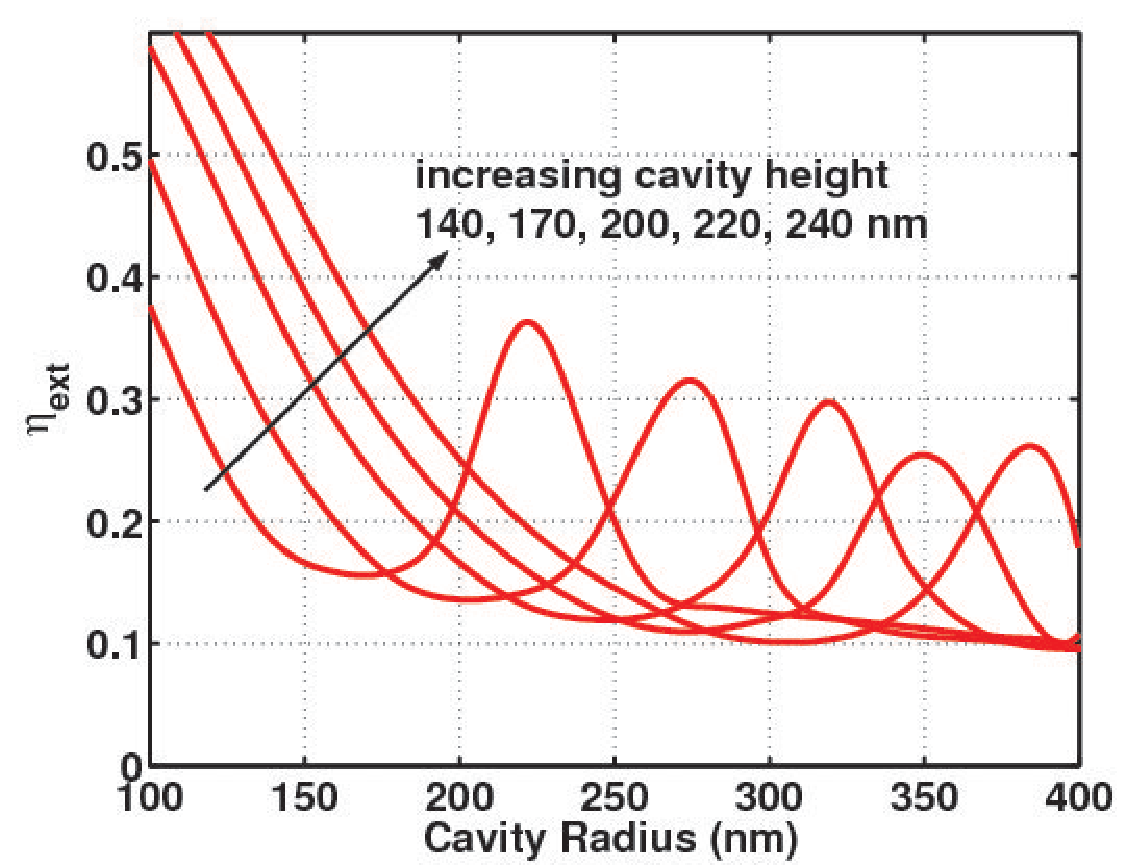}}
\hspace{0.1in}
\subfigure{
\includegraphics[width=3.0in]{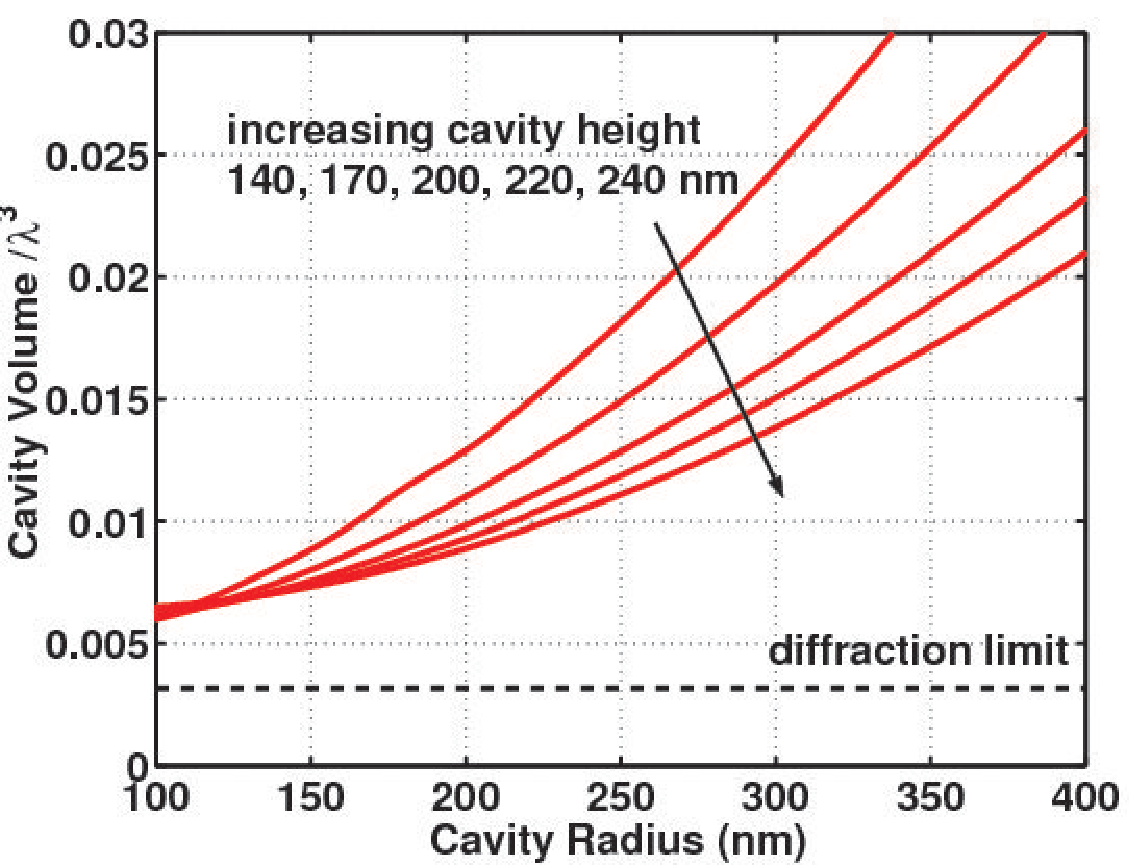}}
\caption{The computed external radiation efficiency $\eta_{ext}$ (LEFT), and the cavity volume (in units of $\lambda^{3}$) (RIGHT), for the TM$_{111}$ modes of circular nanopatch cavities are plotted as a function of the cavity radius for different cavity heights. The cavity heights are 140, 170, 200, 220, and 240 nm. The corresponding mode wavelengths are given in Fig.~\ref{figradwavgamma}. The cavity volume approaches close to the diffraction limit for small cavity radii.}  
\label{figwaveffsize}
\end{figure}

\subsection{Simulation Results for Circular Nanopatch Cavities} \label{secsimresults}
In this Section we present simulation results for the TM$_{111}$ modes of circular nanopatch cavities. In the simulations the cavity height $h$ and the radius $R$ are varied. The thickness of the top and bottom contact/cladding layers is kept fixed and equal to 30 nm each. The contact/cladding layers need to be thick enough to incorporate the depletion regions (without getting fully depleted) at the interface with the smaller bandgap gain medium and also the dipole layers at the interface with the metal layers. The assumed large doping level of $4\times 10^{18}$ 1/cm$^{3}$ for both the n- and p-type contact/cladding layers is therefore necessary. The thickness of the gain region is the cavity height minus the thickness of the two contact/cladding layers. The thickness of the top metal layer is assumed to be 100 nm. The computed fields for the TM$_{111}$ mode are shown in Fig.~\ref{figcirsnl} and show the large confinement of the horizontal component of the field in the gain region. Fig.~\ref{figradwavgamma} shows the wavelengths of the TM$_{111}$ modes as a function of the cavity radius for different cavity heights. Modes with wavelengths in the entire 1.0 $\mu$m to 1.6 $\mu$m range are possible with the cavity dimensions presented in Fig.~\ref{figradwavgamma}. In the 1.0-1.8 $\mu$m wavelength range the interaction between the surface plasmons and the confined optical mode is weak and the mode shape does not deviate significantly from the one given by Eqs.(\ref{eqmode1})-(\ref{eqmode3}). Consequently, a semi-analytical expression for the wavelength of the TM$_{mnp}$ modes can be obtained using the approximate boundary condition that the radial component of the electric field must vanish at the periphery of the cavity (i.e. at $r=R$). If the $n$-th root of the derivative of the $m$-th Bessel function is $\xi_{mn}$, then the wavelength $\lambda_{mnp}$ of the TM$_{mnp}$ mode can be written approximately as,
\begin{equation}
\lambda_{mnp} = \frac{ 2\pi \, \sqrt{\epsilon} }{\sqrt{ (\xi_{mn}/R)^{2} + (p \pi/(h+2\Delta))^{2}}} \label{eqformula}
\end{equation}
Here, $\epsilon$ is the average dielectric constant of the cavity and $\Delta$ is the penetration depth of the field in the top and bottom metal layers. Fig.~\ref{epfit} indicates that the penetration depth is in the 19-21 nm range for free plasmons in the 1.0-1.8 $\mu$m wavelength range. Simulations confirm these numbers in the case of nanopatch cavities as well. Assuming a constant wavelength independent penetration depth of 20 nm, an average cavity index of 3.45, the analytically calculated mode wavelengths (DASHED lines) for the TM$_{111}$ modes are compared with the numerical simulation results (SOLID lines) in Fig.~\ref{figradwavgamma}. The agreement is good except for small radii when the energy in the fringing fields is not small. Fig.~\ref{figradwavgamma} also displays the fraction $\Gamma_{e}$ of the cavity electromagnetic energy confined in the gain region as a function of the cavity radius for different cavity heights and shows that between 55$\%$-85$\%$ of the mode energy is confined in the gain region. 

In plotting the simulated characteristics of the cavity modes one can choose to plot them as a function of the cavity radius or the mode wavelength since there is one-to-one correspondence between the two, as shown in Fig.~\ref{figradwavgamma}. In the plots discussed below, we have chosen the former scheme. The corresponding mode wavelengths can be referenced from Fig.~\ref{figradwavgamma}. Fig.~\ref{figwavtpq} shows the cavity photon lifetimes $\tau_{p}$ and the cavity quality factors $Q$ for the TM$_{111}$ mode. The resonances visible in Fig.\ref{figwavtpq} appear when the diameter of the cavity is slightly smaller than $\lambda/2$ ($\lambda$ is the mode wavelength). Fig.~\ref{figwaveffsize} shows the external radiation efficiency $\eta_{ext}$. For small cavity radii and larger cavity heights the radiative losses dominate, and for large cavity radii the intrinsic losses dominate. In order to obtain decent external efficiencies the radiative losses cannot be much smaller than the intrinsic losses. Therefore, there is little to gained in terms of laser performance by reducing the radiative losses much below the intrinsic losses. An interesting aspect of the nanopatch cavities is their small cavity size. Fig.~\ref{figwaveffsize} shows the cavity volume (not the mode volume) in units of $\lambda^{3}$ as a function of the cavity radius. For comparison, the diffraction limit for the mode volume $(\lambda/2n)^{3}$ is also indicated in the figure (using an average index value of 3.45 for the dielectric part of the cavity). Although the smallest achievable cavity volume using plasmon confinement is neither related to, nor constrained by, the diffraction limit $(\lambda/2n)^{3}$, the diffraction limit has become a useful standard for measuring light confinement in small cavities~\cite{yablan}. Fig.~\ref{figwaveffsize} shows that the cavity volume approaches values close to the diffraction limit for small radii.

\subsection{Far-Field Radiation Patterns of Circular Nanopatch Cavities}
\begin{figure}[tbp]
\centering
\subfigure{
\includegraphics[width=2.5in]{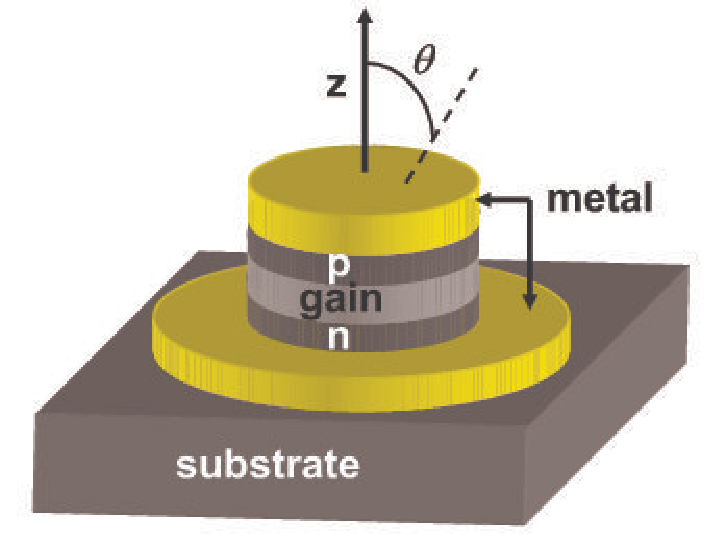}}
\hspace{0.0in}
\subfigure{
\includegraphics[width=3.0in]{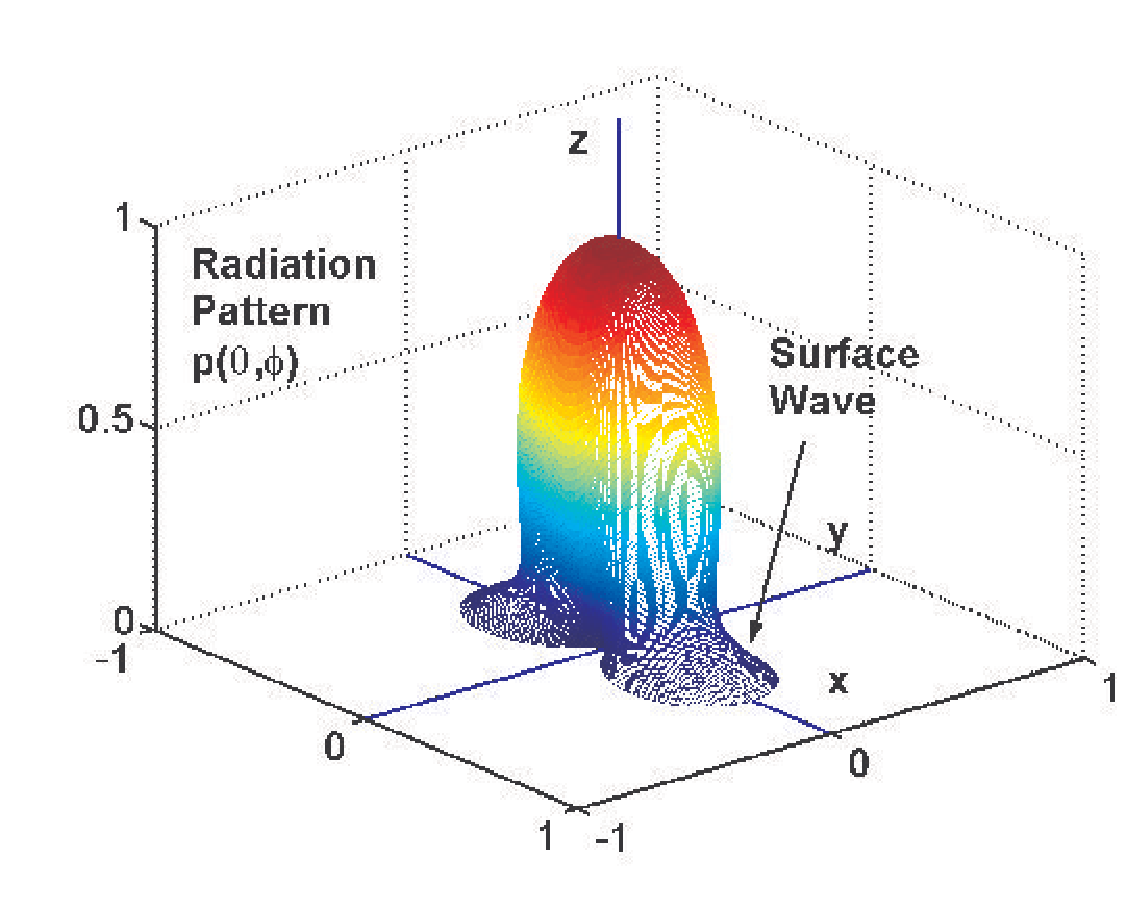}}
\caption{A circular nanopatch laser (LEFT) and the computed far-field radiation pattern (RIGHT) for the TM$_{111}$ mode of a 160 nm radius and 220 nm cavity height laser are shown. The angular dependence of the $z$-component of the field is assumed to be proportional to $\cos(\phi)$ rather than $\exp(\pm i\, \phi)$. A weak surface plasmon wave emitted along the bottom metal layer is also visible in the radiation pattern.}  
\label{figcirnanorad}
\end{figure}

\begin{figure}[tbp]
\centering
\subfigure{
\includegraphics[width=2.9in]{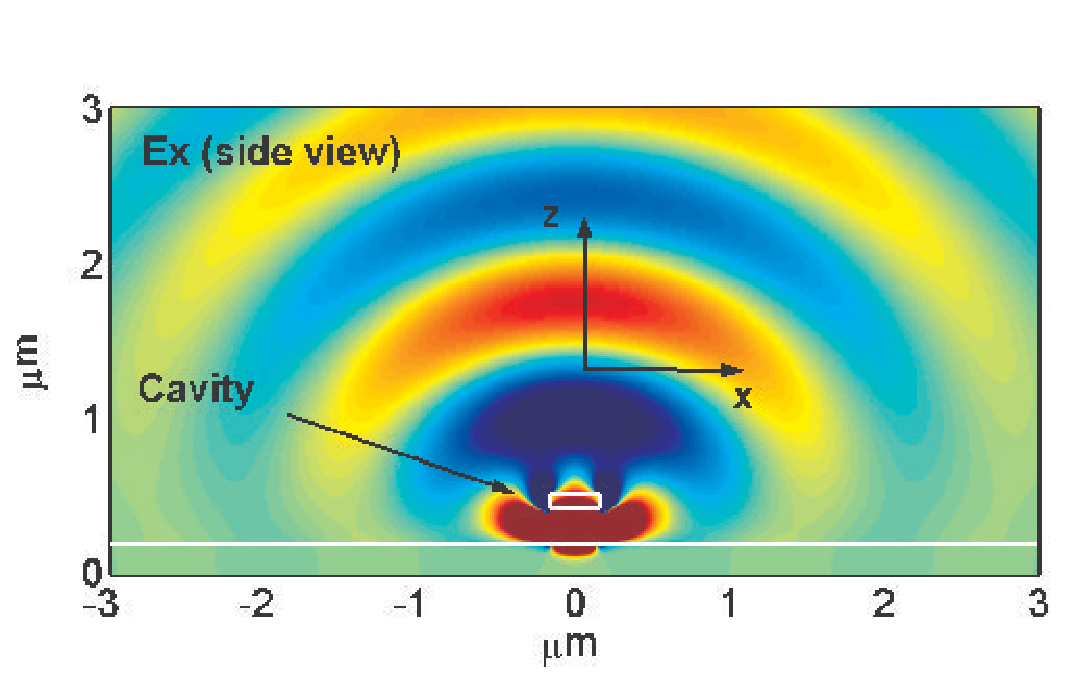}}
\hspace{0.0in}
\subfigure{
\includegraphics[width=2.9in]{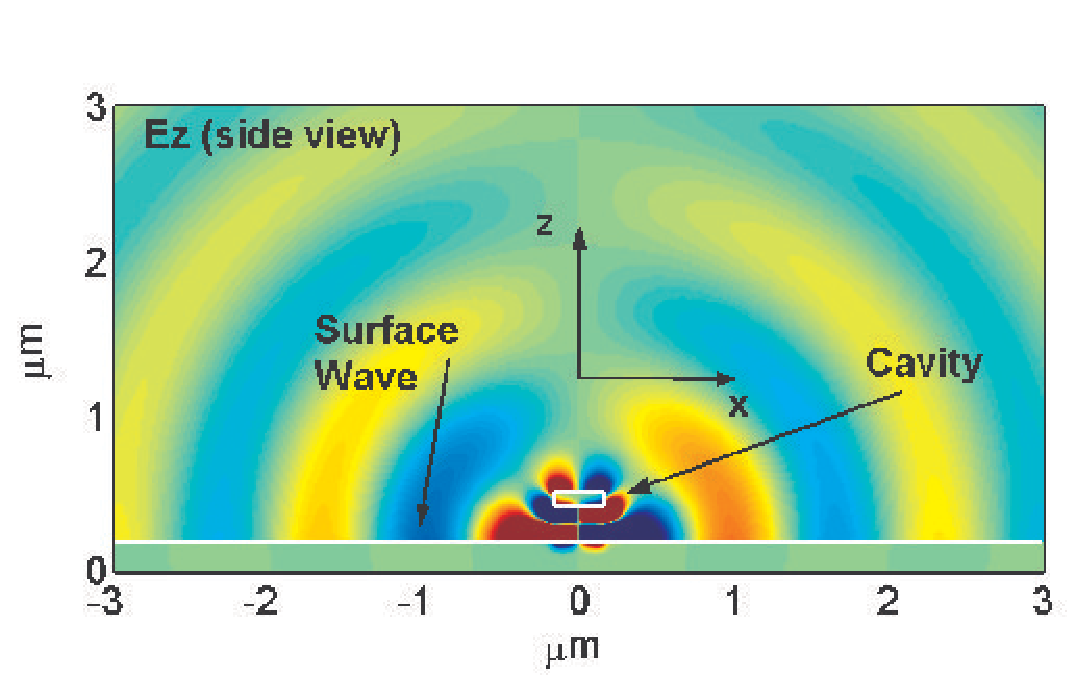}}
\caption{The computed $x$-component (LEFT) and the $z$-component (RIGHT) of the radiated field for the TM$_{111}$ mode of a circular nanopatch cavity of radius 160 nm and height 220 nm are shown.  The angular dependence of the $z$-component of the field is assumed to be proportional to $\cos(\phi)$ rather than $\exp(\pm i\, \phi)$. The radiation fields are shown in the E-plane ($\phi=0$). The contrast in the figures has been greatly reduced to show the radiated fields compared to the much stronger fields inside the cavity.}
\label{figcirrad}
\end{figure}

A useful property of the proposed nanopatch lasers is their surface-normal emission characteristics. Nanopatch lasers can emit in the surface normal direction with a single-lobe far-field radiation pattern and a relatively narrow beam waist. The radiation pattern $p(\theta,\phi)$ is defined as,
\begin{equation}
p(\theta,\phi) = \frac{ \vec{S}(r,\theta,\phi).\hat{r}}{\vec{S}(r,\theta,\phi).\hat{r} |_{{\rm max\,\,over} \,\, \theta , \phi}}
\end{equation}
where $\vec{S}(r,\theta,\phi)$ is the Poynting vector. In simulations, the radiation pattern is found by finding the Poynting vector from the calculated fields over a hemispherical surface of radius large enough such that cavity near-fields do not affect the results. Typically, a value of radius between 4-6 $\mu$m is an adequate compromise between accuracy and limited computational resources. For a circular nanopatch cavity, the radiation pattern will not have a $\phi$-dependence if the modes are assumed to vary with $\phi$ as $\exp(\pm i\,m\,\phi)$. On the other hand, if the $\phi$-dependence of the modes is assumed to be $\cos(m\,\phi)$ or $\sin(m\,\phi)$ then the radiation pattern will also exhibit a $\phi$-dependence. The radiation pattern $p(\theta)$ for $\exp(\pm i\,m\,\phi)$ modes can be obtained by averaging the radiation pattern $p(\theta,\phi)$ for $\cos(m\,\phi)$ or $\sin(m\,\phi)$ modes over $\phi$. Fig.~\ref{figcirnanorad} shows the far-field radiation pattern for the TM$_{111}$ mode of a 160 nm radius and 220 nm cavity height circular nanopatch cavity. The angular dependence of the $z$-component of the field is assumed to be proportional to $\cos(\phi)$ rather than $\exp(\pm i\, \phi)$ to show the $\phi$-dependence of the pattern. The radiation is emitted out from the sides of the cavity but interferes constructively in the upward direction. This is also shown explicitly in Fig.~\ref{figcirrad}. A weak surface plasmon wave is also emitted  along the bottom metal layer. The surface wave decays as it propagates and its power would have made no contribution to the radiation pattern if the radiation pattern were computed over a hemispherical surface of much larger radius.


\subsection{Analytical Model for the Far-Field Radiation Patterns of Nanopatch Cavities}
\begin{figure}[tbp]
\centering
\subfigure{
\includegraphics[width=2.4in]{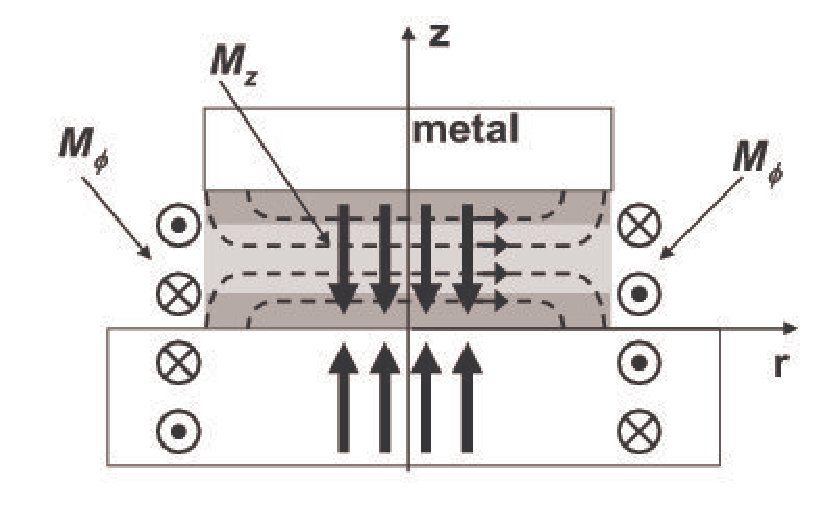}}
\hspace{0.0in}
\subfigure{
\includegraphics[width=1.8in]{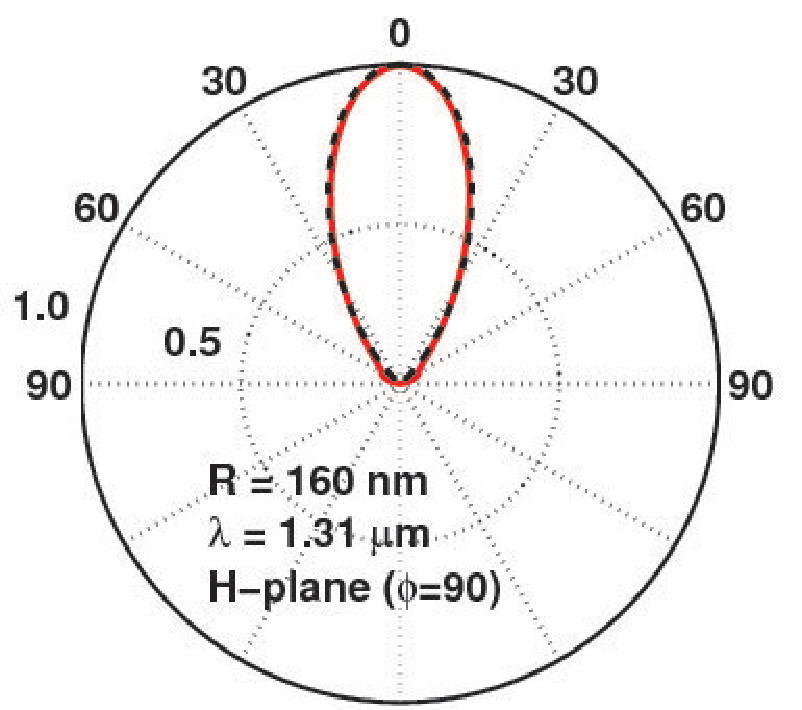}}
\hspace{0.0in}
\subfigure{
\includegraphics[width=1.8in]{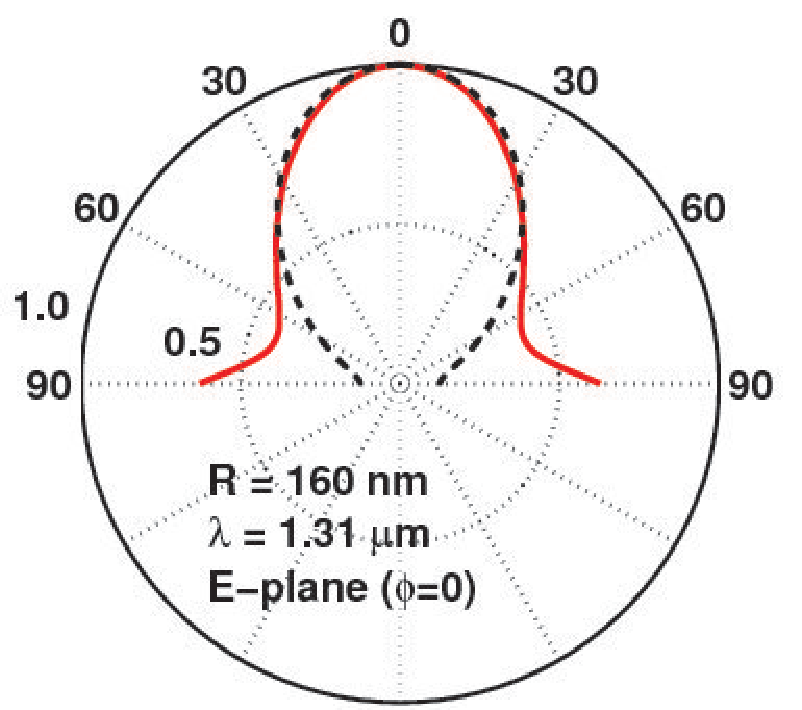}}
\caption{(LEFT) The components of the equivalent magnetic surface current density and their images are depicted for the TM$_{111}$ mode of a circular nanopatch cavity. (MIDDLE and RIGHT) The simulated (SOLID) and the analytically calculated (DASHED) radiation patterns $p(\theta,\phi)$ for the TM$_{111}$ mode of a circular nanopatch cavity are shown in the H-plane ($\phi=90$-degrees) and in the E-plane ($\phi=0$). The cavity height is 220 nm and the cavity radius is 160 nm. The $\phi$-dependence of the $z$-component of the field is assumed to be $\cos(\phi)$. The error for $\theta$ close to 90-degrees in the E-plane is due to the fact that surface plasmon waves are not adequately described by the analytical model based on equivalent magnetic current densities.}
\label{figmagnet}
\end{figure}

The radiation patterns, and the surface-normal emission of radiation in particular, can be understood and estimated from considerations similar to those used in the analysis of aperture antennas at RF/microwave frequencies~\cite{balannis}. The sides of the nanopatch cavities can be considered as apertures from which radiation takes place. In order to calculate the radiated fields, equivalent electric and magnetic surface current densities, $\vec{K}(\vec{r})$ and  $\vec{M}(\vec{r})$, respectively, given by,
\begin{equation}
\vec{K}(\vec{r}) = \hat{n}\times \vec{H}(\vec{r}) |_{\rm boundary}
\end{equation}
\begin{equation}
\vec{M}(\vec{r}) = -\hat{n}\times \vec{E}(\vec{r}) |_{\rm boundary}
\end{equation}
can be assumed to exist at the surface of the cavity and the radiated fields can be assumed to be generated by these current densities~\cite{balannis}. Here, $\hat{n}$ is a unit vector perpendicular to the surface of the cavity. Since the tangential magnetic fields at the sides of the nanopatch cavities are very small (and equal to zero from Eqs.(\ref{eqmode1})-(\ref{eqmode3})), the electric current density can be ignored for simplicity. Unlike at RF/microwave frequencies, the metal layers at near-IR frequencies cannot be considered as perfectly conducting since the real part of the dielectric constant $\epsilon(\omega)$ dominates at near-IR frequencies, as shown in Fig.~\ref{epfit}. However, the large negative values of $\epsilon(\omega)$ at near-IR frequencies imply that the boundary condition of zero tangential electric field at the surface of the metal holds approximately. Therefore, the magnetic surface current density $\vec{M}(\vec{r})$ will have its image in the bottom metal layer. For the TM$_{111}$ mode of a circular nanopatch cavity, the components of the magnetic surface current density and their images are depicted in Fig.~\ref{figmagnet}. The surface normal emission of radiation by nanopatch cavities can be understood as the radiation emitted by the equivalent magnetic surface current density and its image. In the far-field, if the electric field is represented by a vector potential $\vec{F}(\vec{r})$ using the expression,
\begin{equation}
\vec{E}(\vec{r}) = - \frac{\nabla \times \vec{F}(\vec{r})}{\epsilon_{o}}
\end{equation}
then the vector potential corresponding to the magnetic surface current density $\vec{M}(\vec{r})$ can be written as,
\begin{equation}
\vec{F}(\vec{r}) = \frac{\epsilon_{o} \, \exp(i \, k \, r) }{4 \, \pi \, r} \int \int_{\rm boundary} da' \, \, \vec{M}(\vec{r'})  \exp(- i \, \vec{k} . \vec{r'}) 
\end{equation}
Using the expressions for the field components of the TM modes of a circular nanopatch cavity given in Eqs.(\ref{eqmode1})-(\ref{eqmode3}), the far-field radiation pattern $p(\theta, \phi)$ for the TM$_{111}$ mode can be calculated analytically and the approximate result is,
\begin{eqnarray}
p(\theta,\phi) & \propto & \left[ \sin(k\,h\,\cos(\theta)) \, 2\,k\,h\,\cos(\theta) + B\cos(k\,h\,\cos(\theta)/2) \right]^{2} \cos^{2}(\phi) \nonumber \\
          &    +    &  \left[  \left( \sin(k\,h\,\cos(\theta)) \, 2\,k\,h\,\cos(\theta) +  B\cos(k\,h\,\cos(\theta)/2) \right) \, \cos(\theta)  \right. \nonumber \\
          &         & \left. - A \sin(k\,h\,\cos(\theta)) \, \sin^{2}(\theta) \right]^{2}  \sin^{2}(\phi)
\end{eqnarray}
where, $A = 2 \pi^{2} k/(k_{r}^{2}\,h)$, $k=\omega/c$, $h$ is the cavity height, and $k_{r}=\xi_{11}/R$. In deriving the above expression, the $\phi$-dependence of the $z$-component of the field is assumed to be $\cos(\phi)$. The actual TM$_{111}$ mode is not exactly symmetric with respect to a horizontal plane passing through the center of the cavity, as assumed in Eqs.(\ref{eqmode1})-(\ref{eqmode3}). The value of the dimensionless constant $B$ is chosen to model this asymmetry and numerical simulations show that the value of $B$ is in the 0.05 to 0.2 range. The simulated and the analytically calculated (using the expression above) radiation patterns for the TM$_{111}$ mode of a circular nanopatch cavity are shown in Fig.~\ref{figmagnet}. The cavity height is 220 nm, the cavity radius is 160 nm, and the value of $B$ is 0.15. The agreement in both the H-plane ($\phi=90$-degrees) and E-plane ($\phi=0$) is good except for values of $\theta$ close to 90-degrees in the E-plane. The error for values of $\theta$ close to 90-degrees in the E-plane is due to the fact that surface plasmon waves on the bottom metal layer that are emitted in the E-plane are not adequately described by the analytical model based on the equivalent magnetic current densities.

\begin{figure}[tbp]
\centering
\subfigure{
\includegraphics[width=1.8in]{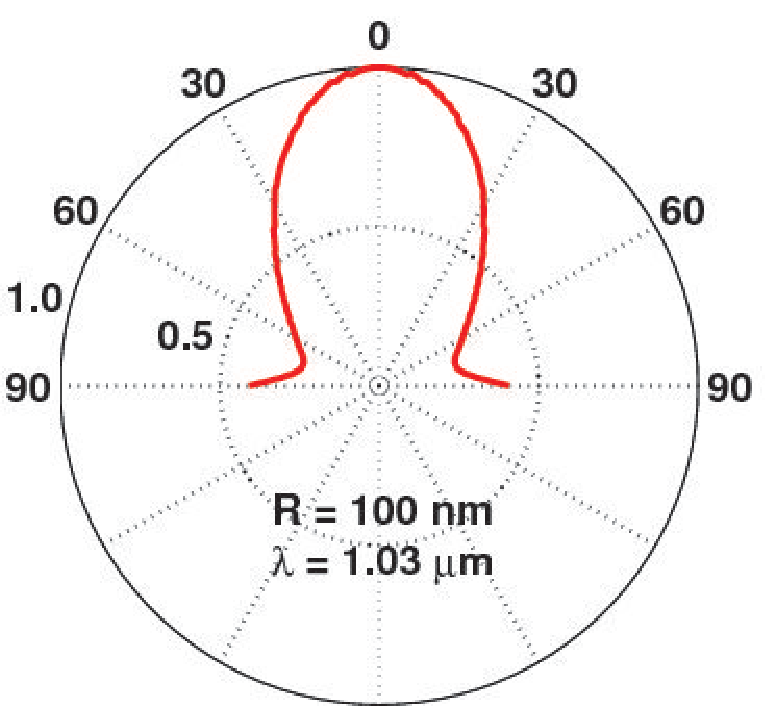}}
\hspace{0.0in}
\subfigure{
\includegraphics[width=1.8in]{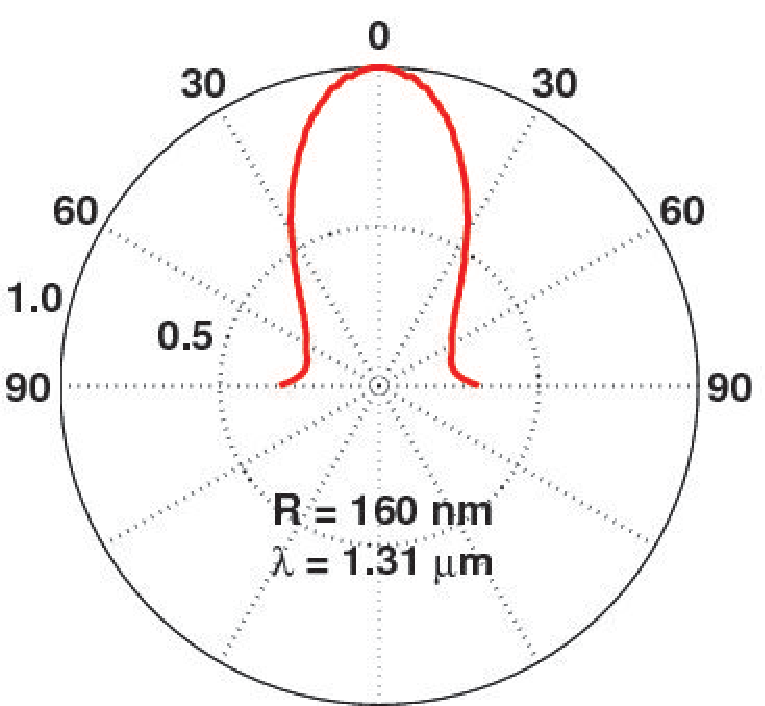}}
\hspace{0.0in}
\subfigure{
\includegraphics[width=1.8in]{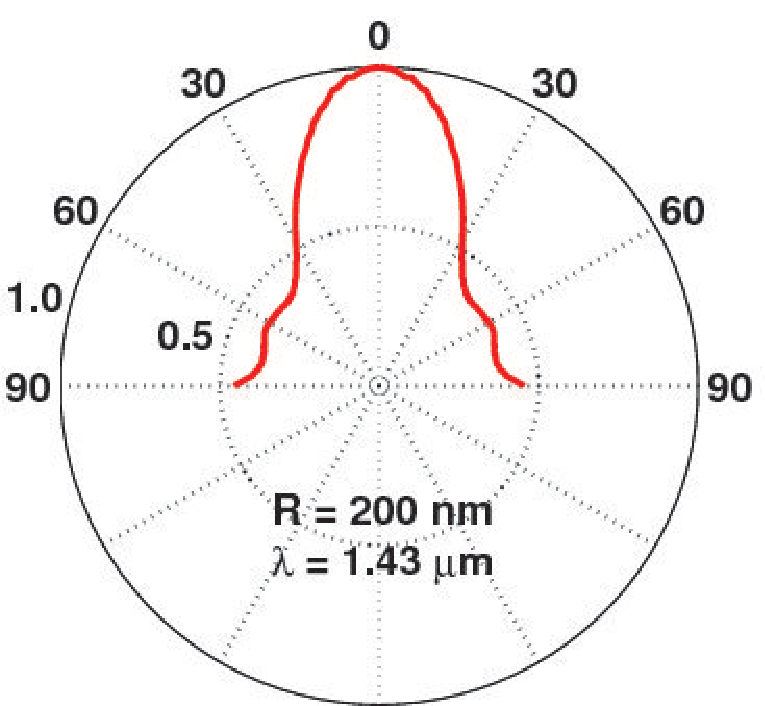}}
\vspace{0.0in}
\subfigure{
\includegraphics[width=1.8in]{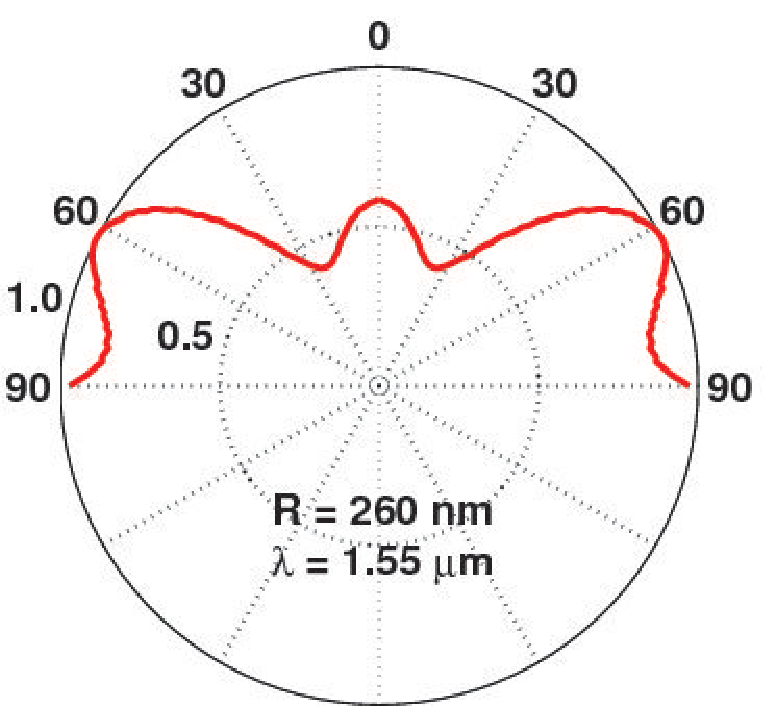}}
\hspace{0.0in}
\subfigure{
\includegraphics[width=1.8in]{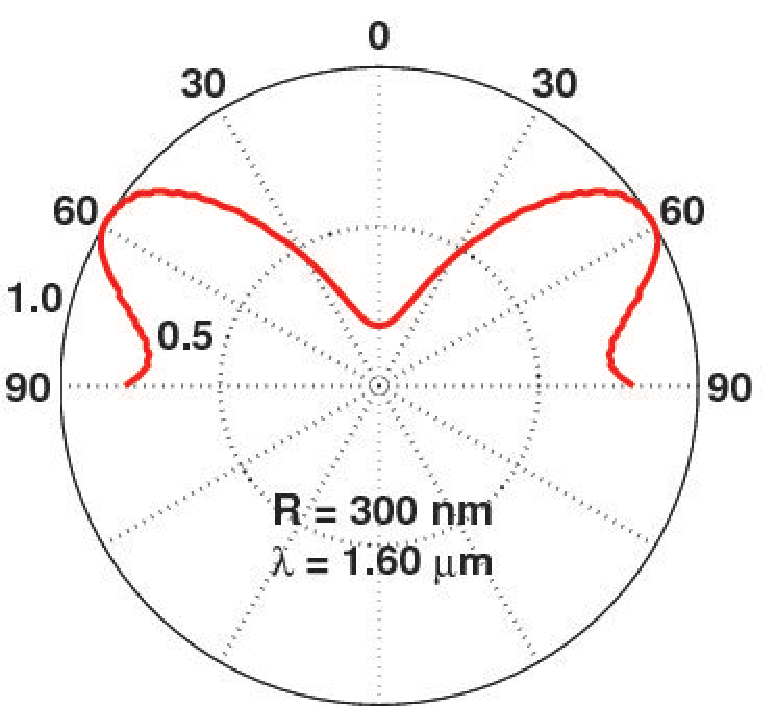}}
\hspace{0.0in}
\subfigure{
\includegraphics[width=1.80in]{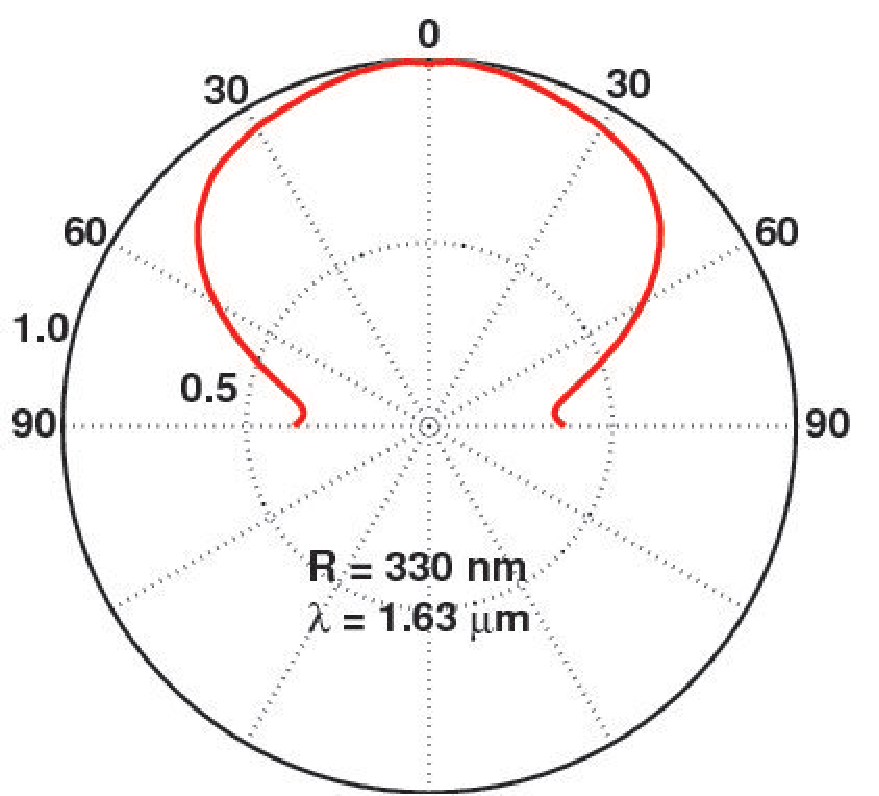}}
\caption{The simulated far-field radiation patterns $p(\theta)$ for the TM$_{111}$ modes of circular nanopatch cavities of height 220 nm and different radii are shown as a function of the angle $\theta$. The $\phi$-dependence of the modes is assumed to be $\exp(\pm i\, \phi)$ and therefore the radiation patterns depend only on the angle $\theta$.}
\label{figcirpat}
\end{figure}

Fig.~\ref{figcirpat} shows the computed far-field radiation patterns $p(\theta)$ for the TM$_{111}$ modes of circular nanopatch cavities of height 220 nm for different cavity radii. The $\phi$-dependence of the modes is assumed to be $\exp(\pm i\, \phi)$ for simplicity. It can be seen that the radiation patterns depend sensitively on the cavity dimensions. In the surface normal direction the radiation is entirely due to the $\phi$-component of the magnetic current density. For small cavity radii ($R<h$ nm), the radiation is dominated by the $\phi$-component of the magnetic current density. For larger cavity radii ($R>h$ nm), the $z$-component of the magnetic current density also contributes strongly to the radiation.

\begin{figure}[tbp]
\centering
\subfigure{
\includegraphics[width=2.0in,angle=-0]{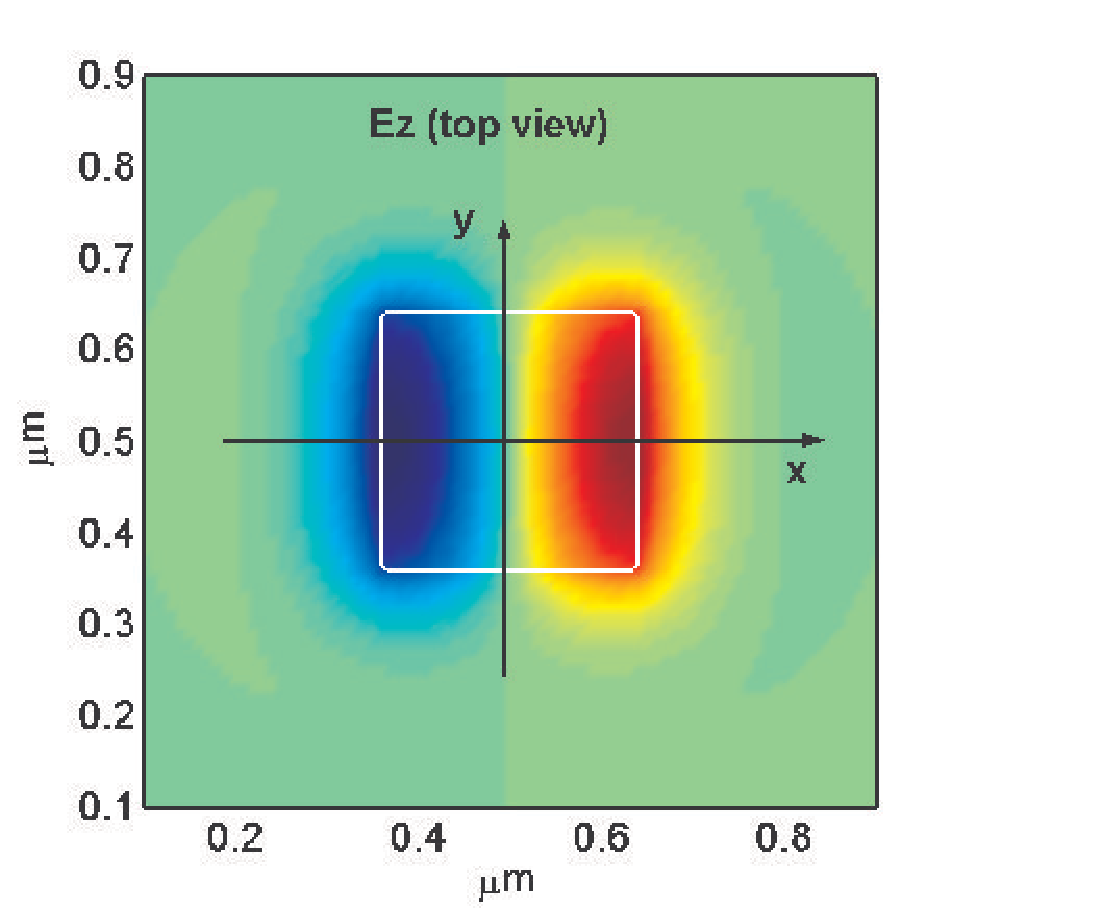}}    
\hspace{0.0in}
\subfigure{
\includegraphics[width=2.0in,angle=-0]{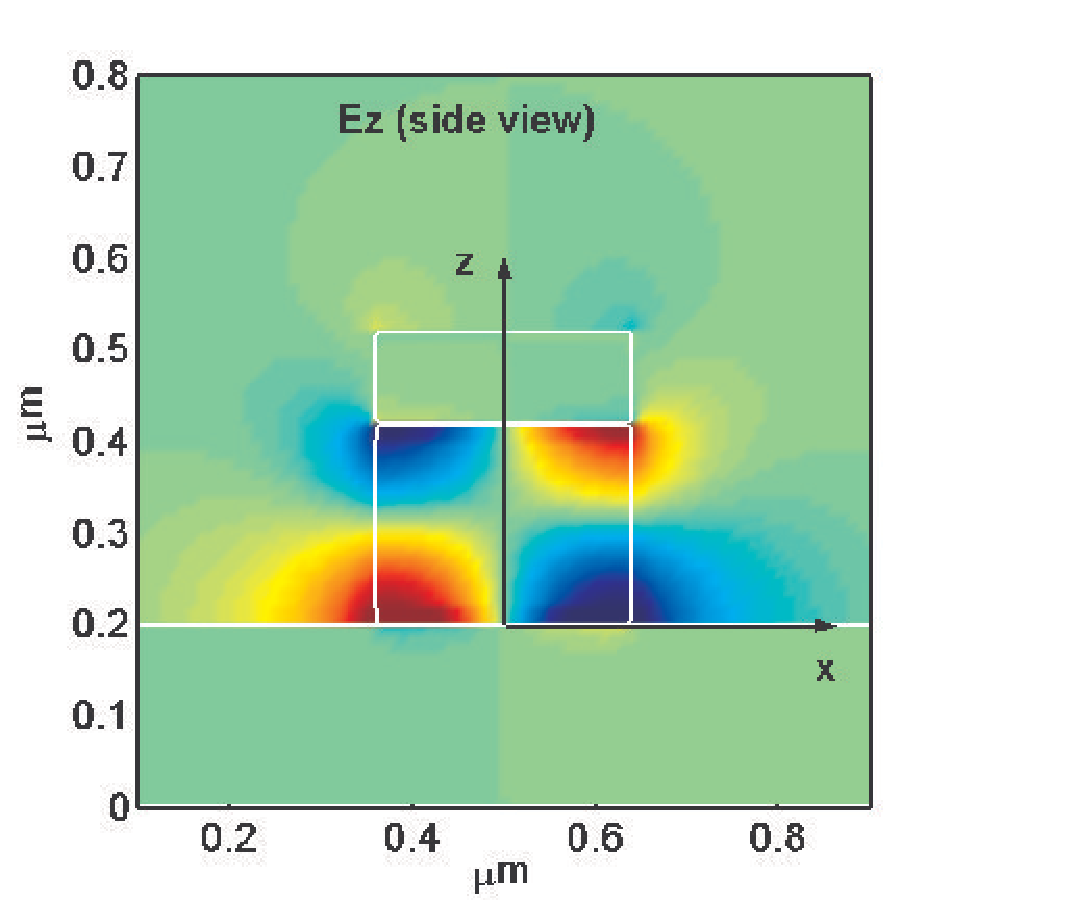}}    
\hspace{0.0in}
\subfigure{
\includegraphics[width=2.0in,angle=-0]{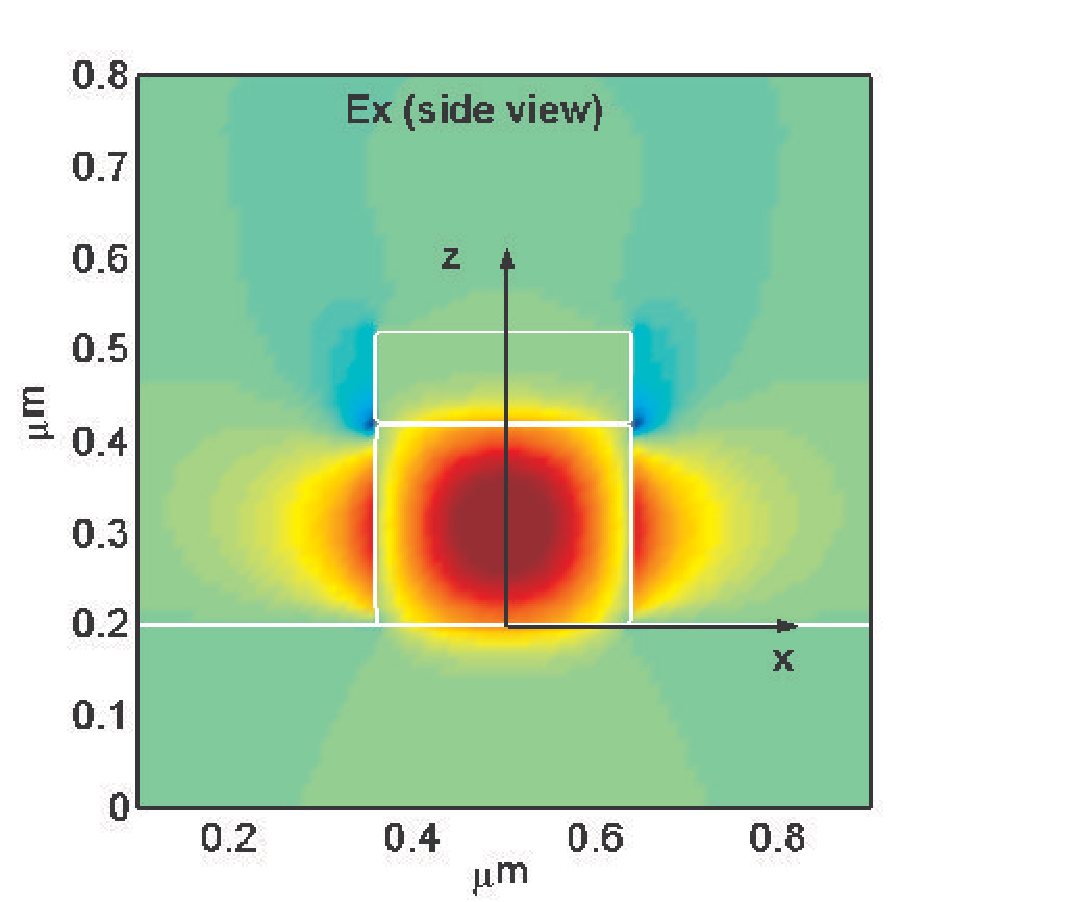}}
\caption{The computed $z$-component(top view) (LEFT), $z$-component(side view) (MIDDLE), and  the $x$-component(side view) of the electric field for the TM$_{111}$-like mode of a square nanopatch cavity of side 280 nm and height 220 nm are shown. The $z$-component (top view) is shown in a horizontal plane just below the top metal. The contrast has been adjusted to show the weak fringing fields.}
\label{figsqsnl}
\end{figure}

\begin{figure}[tbp]
\centering
\subfigure{
\includegraphics[width=2.0in,angle=-0]{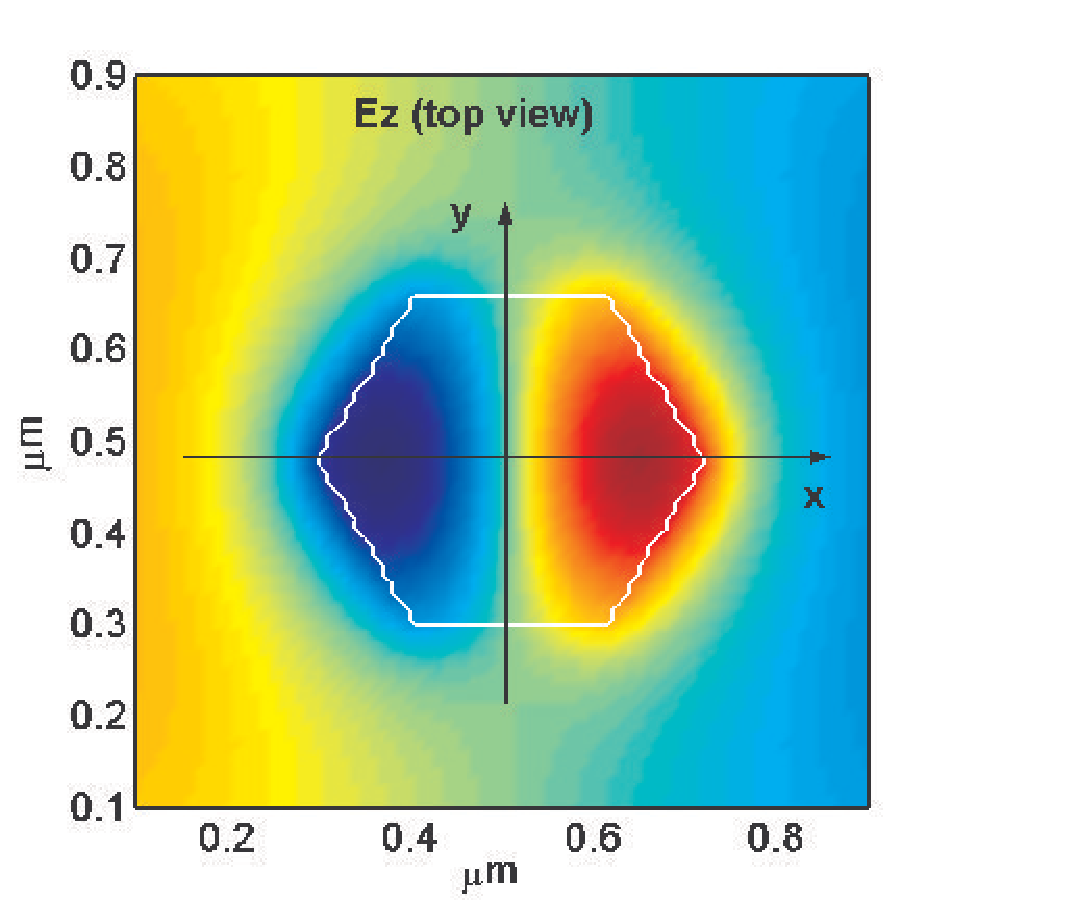}}    
\hspace{0.0in}
\subfigure{
\includegraphics[width=2.0in,angle=-0]{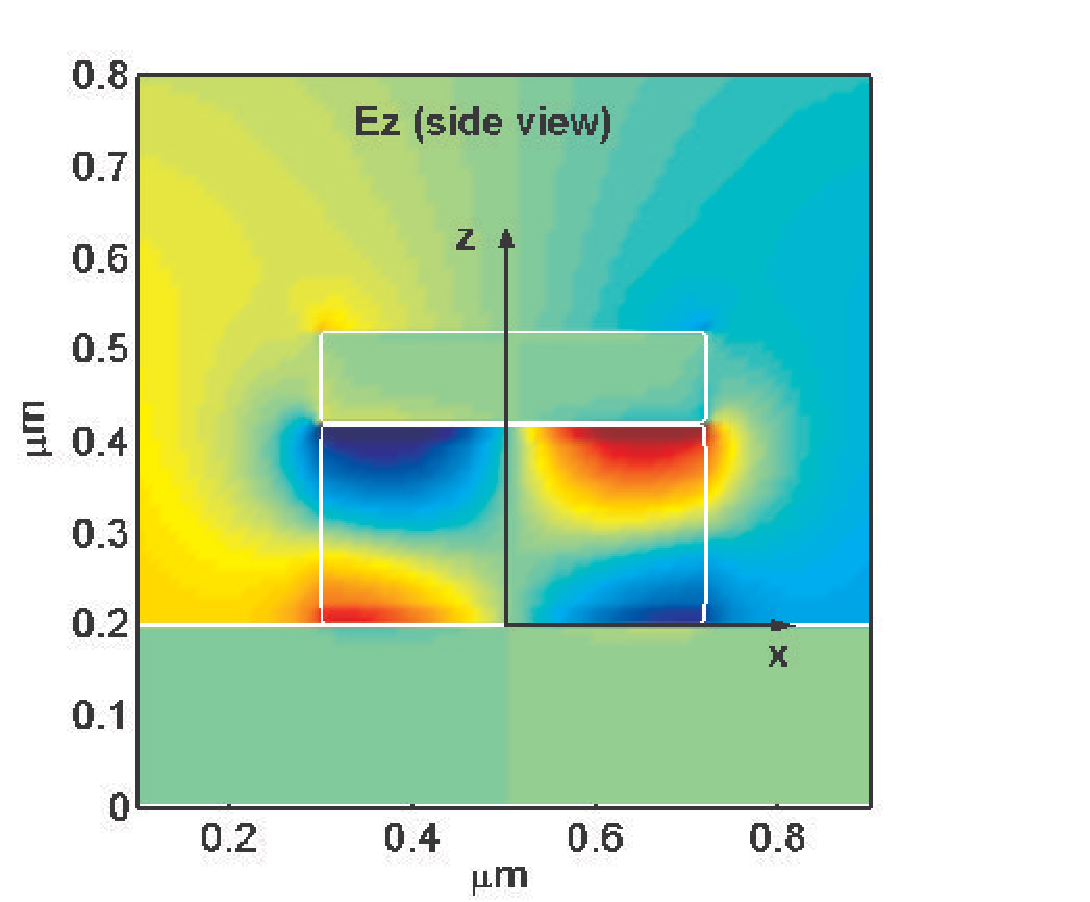}}    
\hspace{0.0in}
\subfigure{
\includegraphics[width=2.0in,angle=-0]{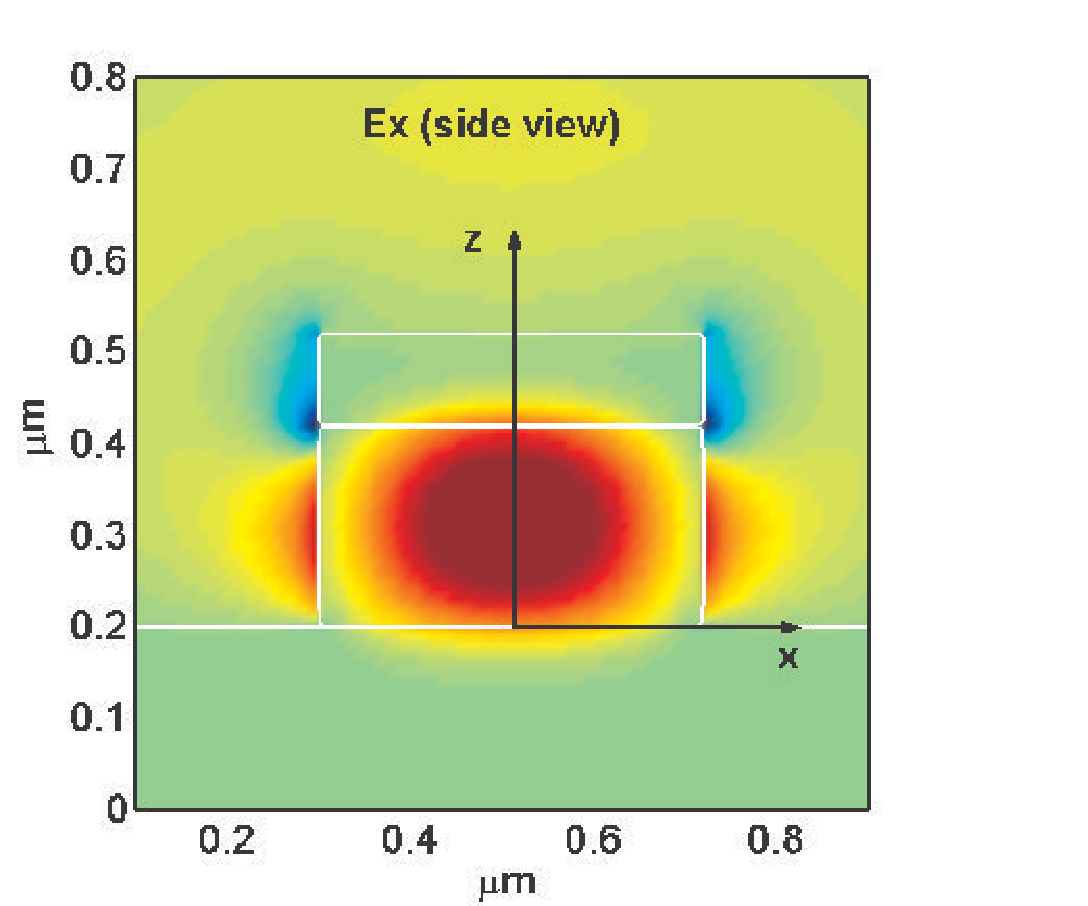}}
\caption{The computed $z$-component(top view) (LEFT), $z$-component(side view) (MIDDLE), and  the $x$-component(side view) of the electric field for the TM$_{111}$-like mode of a hexagonal nanopatch cavity of height 220 nm and side 200 nm are shown. The $z$-component (top view) is shown in a horizontal plane just below the top metal. The contrast has been adjusted to show the weak fringing fields.}
\label{fighexsnl}
\end{figure}

\begin{figure}[tbp]
\centering
\subfigure{
\includegraphics[width=3.0in,angle=-0]{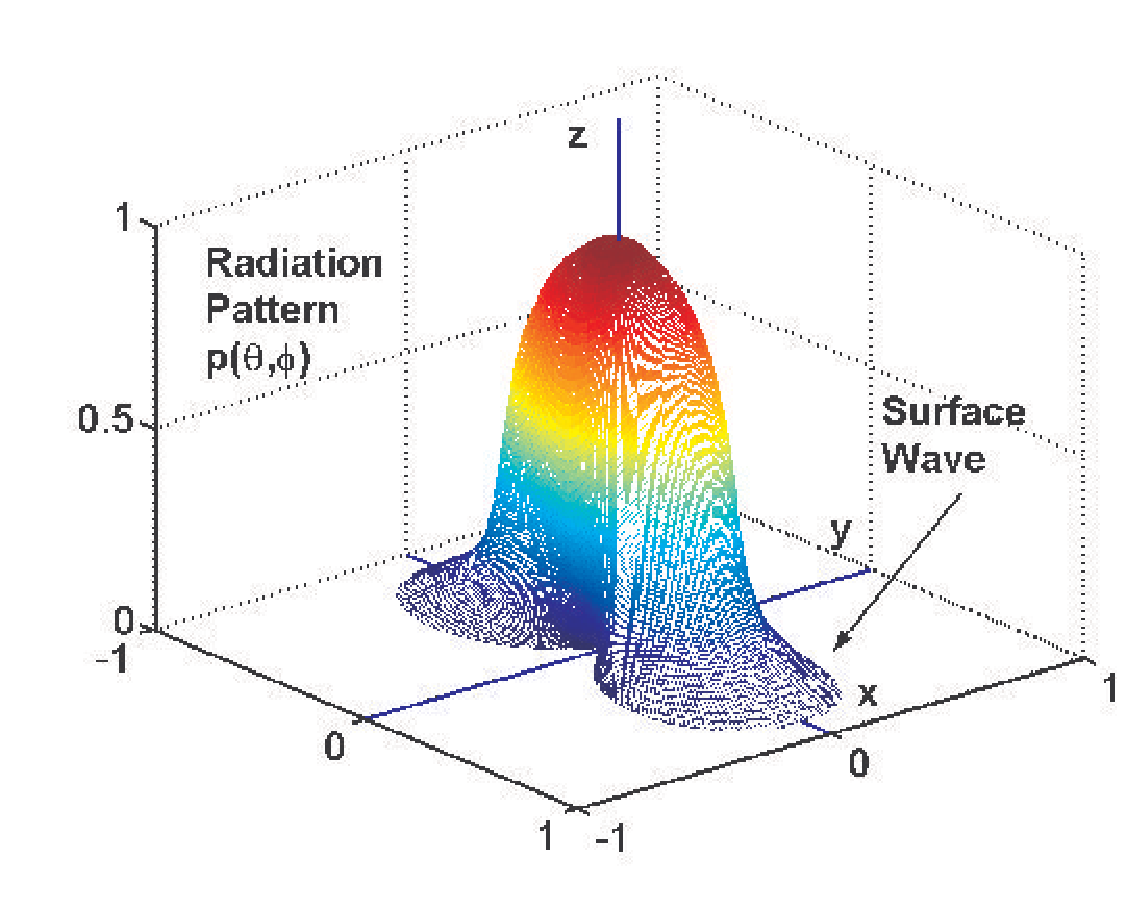}}    
\hspace{0.0in}
\subfigure{
\includegraphics[width=3.0in,angle=-0]{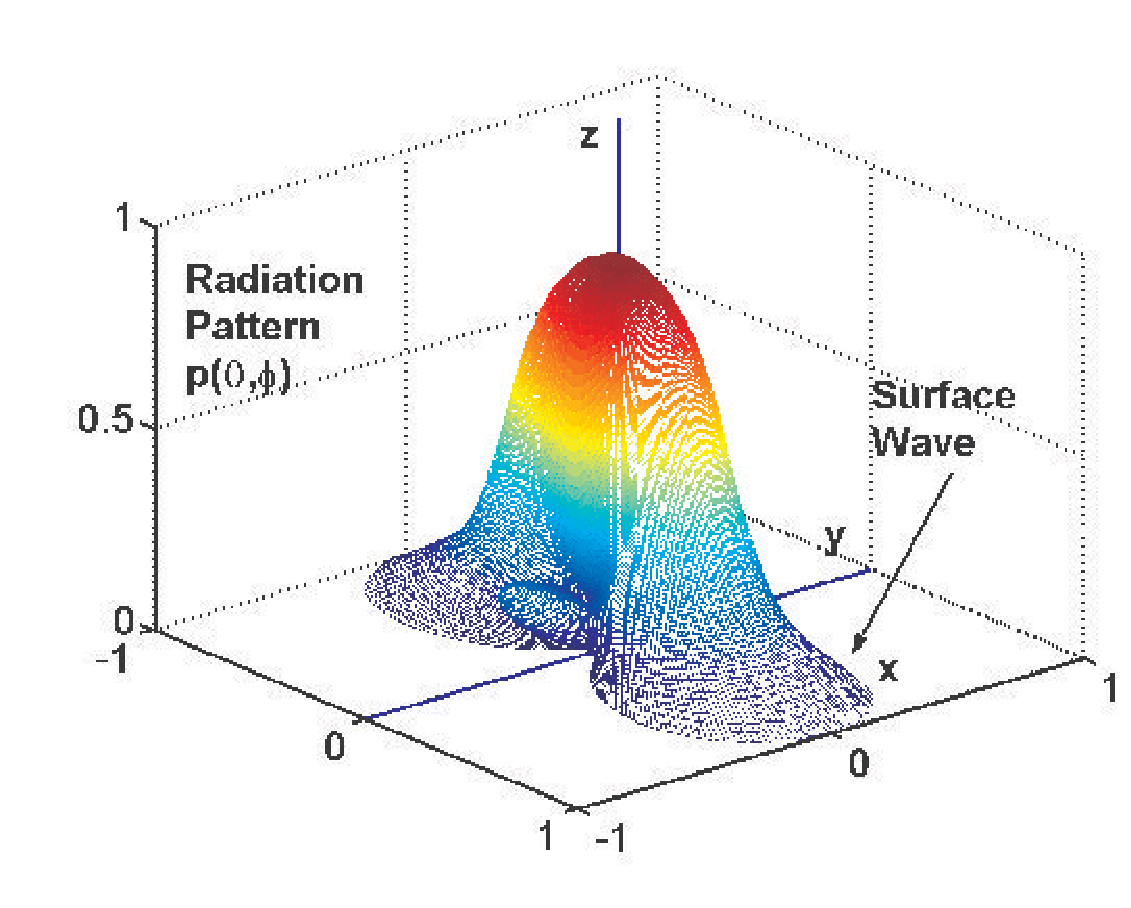}}    
\caption{The computed far-field radiation patterns for the TM$_{111}$-like modes of a square (LEFT) and hexagonal (RIGHT) nanopatch cavities are shown. The cavity dimensions are the same as in Figs.~\ref{figsqsnl} and \ref{fighexsnl}.}
\label{figradsqhex}
\end{figure}

\subsection{Modes, Degeneracies, and Far-Field Radiation Patterns in Square and Hexagonal Nanopatch Cavities}
In actual fabricated circular nanopatch cavities the angular symmetry is expected to be broken unintentionally and the modes are expected to have $\cos(m\,\phi)$ or $\sin(m\,\phi)$ dependence rather than $\exp(\pm i \, m\phi)$ dependence. The angular symmetry can be intentionally broken using the square and hexagonal nanopatch cavities that also support modes that resemble the TM$_{111}$ modes of circular nanopatch cavities. These modes have a horizontal field component that is symmetric with respect to a horizontal plane passing through the center of the cavity and a vertical field component that is antisymmetric with respect to the same plane. The antisymmetry of the vertical component of the field at the cavity edges helps in reducing the radiation losses and the horizontal field component enjoys a large overlap with the gain region. 

The calculated fields for a square nanopatch cavity of height 220 nm and side 280 nm and a hexagonal nanopatch cavity of height 220 nm and side 200 nm are shown in Figs.~\ref{figsqsnl} and \ref{fighexsnl}, respectively. The mode wavelength and the cavity-$Q$ for the square cavity are 1.30 $\mu$m and 196, respectively, and 1.41 $\mu$m and 242, respectively, for the hexagonal cavity. These cavity-$Q$ values are comparable to those of circular nanopatch cavities. The continuous angular symmetry in circular nanopatch cavities implies that the two modes with the same angular index $m$ (for $m \ne 0$) are degenerate. The symmetries in case of square and hexagonal nanopatch cavities are discrete, and correspond to the $C_{4v}$ and $C_{6v}$ point groups, respectively~\cite{skoda}. The character tables of the $C_{4v}$ and $C_{6v}$ point groups are given in Appendix~\ref{app1}. The modes for the square and hexagonal nanopatch cavities shown in Figs.~\ref{figsqsnl} and \ref{fighexsnl} correspond to the two-dimensional $E$ and $E_{1}$ representations of the $C_{4v}$ and $C_{6v}$ point groups, respectively, and are therefore each doubly degenerate. Most of the properties of square and hexagonal nanopatch cavities are not too different from those of circular nanopatch cavities. The far-field radiation patterns for a square nanopatch cavity of height 220 nm and side 280 nm and a hexagonal nanopatch cavity of height 220 nm and side 200 nm are shown in Fig.~\ref{figradsqhex}, and resemble the radiation patterns of circular nanopatch cavities (see Fig.~\ref{figcirnanorad}).

\section{Threshold Gain and Threshold Current of Nanopatch Lasers} \label{secgain}

\subsection{Threshold Gain}
\begin{figure}[tbp]
\centering
\subfigure{
\includegraphics[width=3.0in]{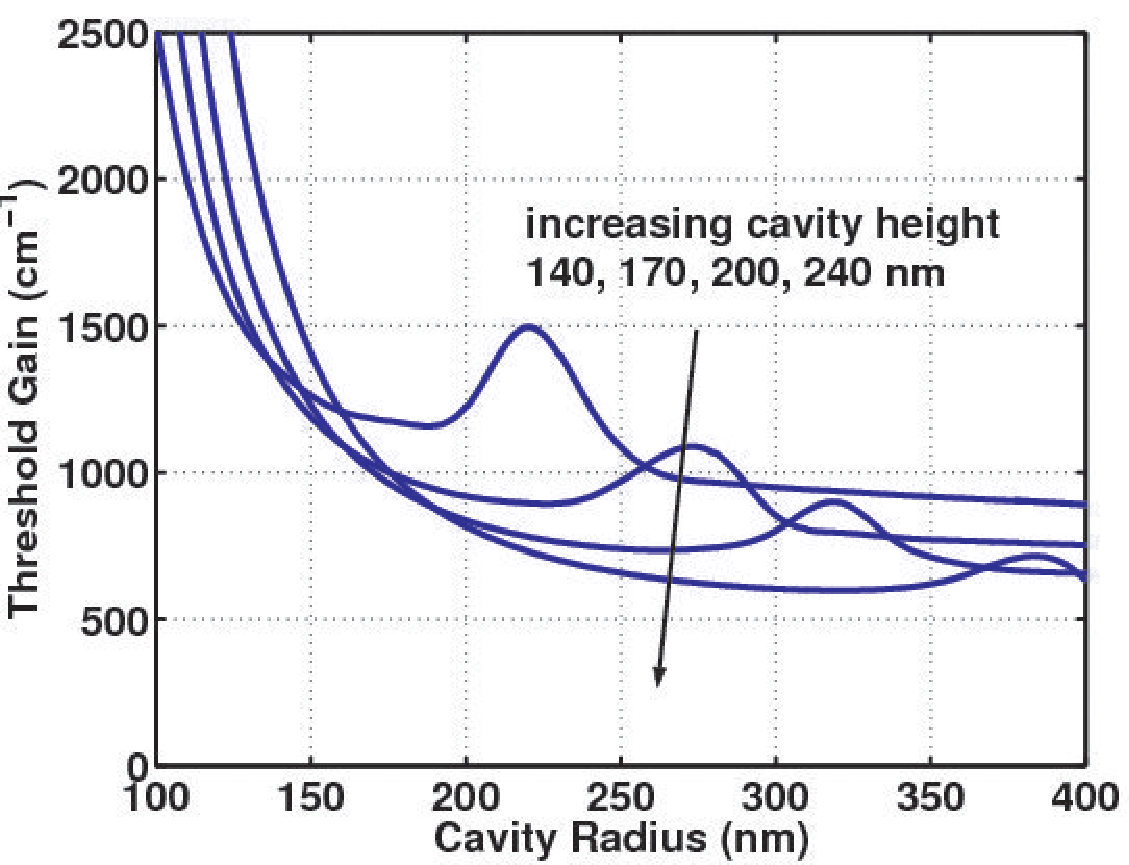}}
\hspace{0.0in}
\subfigure{
\includegraphics[width=2.9in]{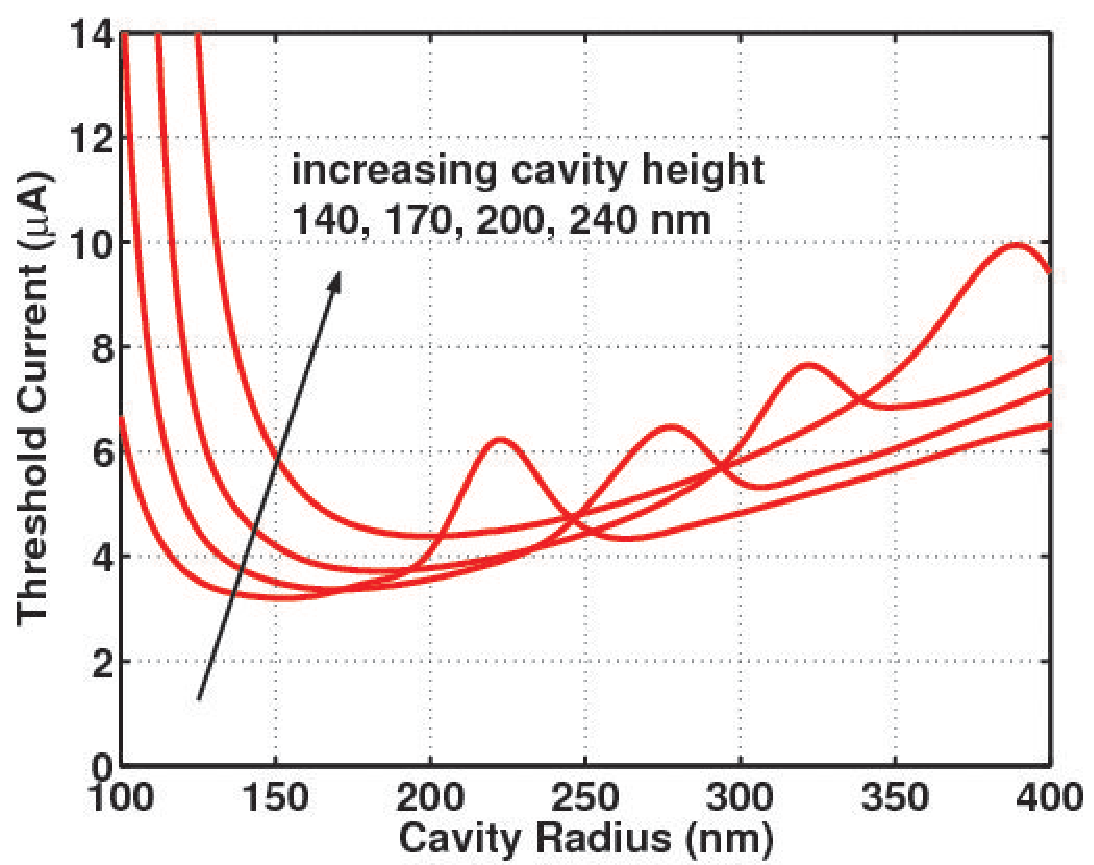}}
\caption{(LEFT) The computed material threshold gain $g_{th}$ for the TM$_{111}$ mode of a circular nanopatch laser for different cavity heights are shown. (RIGHT) The threshold current $I_{th}$ for a circular nanopatch laser are shown as a function of the cavity radius for different cavity heights. The cavity heights are 140, 170, 200, and 240 nm. The top and bottom contact/cladding layers are 30 nm thick each.  The values of various parameters used in the calculations are listed in Table.1.}
\label{figgain}
\end{figure}

In a semiconductor laser the material threshold gain $g_{th}$, the mode energy gain confinement factor $\Gamma_{e}$, the gain medium refractive index $n$, and the photon lifetime $\tau_{p}$ are related as~\cite{coldren},
\begin{equation}
\Gamma_{e} \, \frac{c}{n} \, g_{th} = \frac{1}{\tau_{p}} \label{eqthreshold}
\end{equation}
where, $c$ is the speed of light in free space. The strong mode confinement in the gain region provided by the surface plasmons in nanopatch cavities results in large modal gain values that are needed to compensate for the large cavity losses. The photon lifetimes and the mode energy gain confinement factors in conventional edge-emitting semiconductor lasers typically range from 1-4 ps and 0.04-0.08$\%$, respectively~\cite{coldren}. The results presented in Section~\ref{secsimresults} show that the photon lifetimes in nanopatch cavities are almost 10 times smaller than in conventional semiconductor lasers but the mode energy gain confinement factors are almost 10 times larger compared to their values in conventional semiconductor lasers. Therefore, the material threshold gain values in nanopatch lasers are not expected to be much different than in conventional semiconductor lasers. Although Eq.~\ref{eqthreshold} can be used to find the material threshold gain values from the calculated values of $\Gamma_{e}$ and $\tau_{p}$, this procedure does not take into account modification of the mode profile in the presence of material gain. A more accurate method to obtain the threshold gain is to calculate the complex mode frequency in the presence of the material gain using the techniques discussed in Section~\ref{sectech} and then iteratively adjust the value of the material gain until the imaginary part of the complex mode frequency becomes zero (or sufficiently small). This procedure has been used this paper and the results for the threshold gain of circular nanopatch lasers are shown in Fig.~\ref{figgain}. The lasing mode is assumed to be TM$_{111}$. The threshold gain values are less than 1000 cm$^{-1}$ for cavity radii larger than 175 nm and cavity heights larger than 170 nm. In the 1.5-1.6 $\mu$m wavelength range, gain values smaller than 700 cm$^{-1}$ are achievable. III-V semiconductor materials, such as InGaAsP and InGaAlAs, can easily provide material gain values in the 1200-1600 cm$^{-1}$ range in the 1.2-1.6 $\mu$m wavelength range and are therefore suitable for nanopatch lasers~\cite{coldren}. 

\begin{table}[tbp] 
\caption{Laser parameter values used in calculation of threshold currents~\cite{coldren}}
\begin{center}
\begin{tabular}{|l|l|}
\hline
Parameter & Value \\
\hline
Gain coefficient $g_{o}$ & 2000 cm$^{-1}$ \\
Transparency carrier density $N_{tr}$ & $1.8 \times 10^{18}$ cm$^{-3}$ \\
Auger Coefficient $C$ & $3 \times 10^{-29}$ cm$^{6}$/s \\
Spontaneous emission factor $n_{sp}$ & 1.5 \\
Surface recombination velocity $v_{surf}$ ~\cite{ptho} &  $10^{3}$ cm/s \\
Index of active region $n$ & 3.5 \\ 
\hline 
\end{tabular}
\end{center}
\end{table}

\subsection{Threshold Current}
In this Section, estimates of the threshold currents in semiconductor nanopatch lasers are presented. The threshold carrier density $N_{th}$ can be estimated from the expression for the material gain as a function of the carrier density $N$~\cite{coldren},
\begin{equation}
g(N) = g_{o} \log\left(\frac{N}{N_{tr}}\right)
\end{equation}
The threshold current $I_{th}$ can be related to the electron-hole recombination rate $R(N)$ in the active region at threshold,
\begin{equation}
\frac{I_{th}}{q V} = R(N_{th})
\end{equation}
Here, $q$ is the electron charge and $V$ is the volume of the active region. The carrier density dependent recombination rate $R(N)$ (units: cm$^{-3}$-s$^{-1}$) is assumed to have dominant contributions from surface recombination, spontaneous emission, and Auger scattering. Surface recombination is expected to become more important as the cavity radius decreases and the surface area to volume ratio of the active region increases. The net recombination rate $R(N)$ can be written as,
\begin{eqnarray}
R(N) & = & R_{sp}(N) + R_{auger}(N) + R_{surf}(N) \nonumber \\
  & = & \Gamma_{e} \, \frac{c}{n} \, g(N) \, \frac{n_{sp}}{V_{p}} + C\,N^{3} + v_{surf} \, N \, \frac{A}{V} 
\end{eqnarray}
where, $c$ is the speed of light, $n$ is the index of the active region, $n_{sp}$ is the spontaneous emission factor, $V_{P}$ is the mode volume and is approximately equal to the cavity volume for nanopatch lasers, $C$ is the Auger coefficient, and $v_{surf}$ is the surface recombination velocity and equals $\sim 10^{3}$ cm/s for passivated InGaAsP active layers~\cite{ptho}. $(A/V)$ is the surface area to volume ratio for the active region and equals $2/R$ for a circular nanopatch laser of radius $R$. Given the small cavity size, the spontaneous emission is assumed to occur only in the lasing mode. Using the parameter values given in Table.1, the calculated threshold currents of circular nanopatch lasers are shown in Fig.~\ref{figgain} as a function of the cavity radius for different cavity heights. The parameter values are assumed to be constant whereas in reality they will change slightly as the composition of the gain medium is changed. The parameter values shown in Table.1 correspond to bulk InGaAsP active medium lattice matched to InP and with a bandgap corresponding to 1.3 $\mu$m wavelength~\cite{coldren}. The threshold currents are large for very small radii because the photon lifetimes are small. For large cavity radii the threshold currents are again large because the active region volumes are large. Therefore, there is an optimum value of the cavity radius that results in the minimum value of the threshold current. The minimum values of the threshold current are in the $3-5$ $\mu$A range and correspond to threshold current densities in the 2-3 kA/cm$^{2}$ range.

\section{Conclusion} \label{secfinal}
Challenges associated with the fabrication of nanopatch lasers are not expected to be impossible to meet. Substrate removal techniques used in the fabrication of dual-metal-waveguide semiconductor far-IR lasers can be employed for the realization of the dual-metal nanopatch laser structures~\cite{williams}. However, in the case of nanopatch lasers the semiconductor films obtained after substrate removal would be much thinner and would require more care. The metal used for the top and bottom layers must make good ohmic contacts to the contact/cladding layers, must not react with or diffuse into the semiconductor, and must also have small surface plasmon losses. Silver seems to be the ideal choice. The work function of Silver is very close to that of Titanium and Titanium/Gold ohmic contacts are commonly used for both n- and p-type III-V semiconductors~\cite{ohmic}. Good quality Silver ohmic contacts, which do not require post-deposition rapid thermal anneal, to moderately doped InP have been demonstrated~\cite{silver}. Electrical contact leads to the top metal layer would need to be microfabricated since the laser structure is too small to be contacted directly with electrical probes. Finally, the exposed side walls of the cavity would need to be passivated to reduce surface recombination. Quantum well gain media can also be used in place of bulk gain media in nanopatch lasers. Quantum wells can generally provide larger material gain compared to bulk for the same carrier density due to the reduced density of states~\cite{coldren}. However, a quantum well gain medium would have smaller overlap with the cavity mode and the material gain would also be polarization dependent. The effect of interface roughness on the cavity quality factors has not been considered in this work. In nanoscale lasers, the surface area to volume ratio increases as the cavity dimensions shrink and therefore interface roughness will likely play an important role.   

The ideas presented in this paper show that in the 1.0-1.6 $\mu$m wavelength range surface plasmons can be used to confine light in nanoscale optical cavities. At these longer wavelengths since the surface plasmon fields are not strongly confined to the surface of the metal, they can have significant overlap with an external gain medium. In fact, surface plasmon confinement in the 1.0-1.6 $\mu$m wavelength range can be used to obtain mode gain confinement factors that are substantially larger than in conventional semiconductor lasers thereby providing large modal gain values necessary to achieve lasing.  

The benefits of nanopatch laser structures are not expected to be limited to just reduced dimensions. The small photon lifetimes in nanopatch lasers compared to conventional semiconductor lasers could enable ultrawide bandwidths ($>$ 100 GHz) for direct current modulation. The dual-metal layer structure of nanopatch lasers could enable them to be used as functional elements and placed on desired substrates, such as Silicon microchips, to provide {\it lasers on demand} capabilities for multifunctional micro- and nano-systems. 

The authors would like to thank Dr. Filbert Bartoli for support through the National Science Foundation faculty CAREER award and Dr. Leda Lunardi for support through the National Science Foundation award $\#$0636593, and also ILX lightwave Corp., Intel Inc., and Infotonics Inc. for support.   

\newpage

\appendices

\section{Character Tables for the $C_{4v}$ and $C_{6v}$ Point Groups} \label{app1}
The character tables for $C_{4v}$ and $C_{6v}$ point groups are shown below. The notations for the group symmetry operations are according to Ref.~\cite{skoda}.\\

\begin{tabular}{|c | c c c c c|}
\hline
\multicolumn{6}{|c|}{$C_{4v}$ Point Group} \\
\hline
$C_{4v}$ & $E$ & $2C_{4}$ & $C_{2}$ & $2\sigma_{v}$ & $2\sigma_{d}$ \\
\hline
$A_{1}$ & 1 & 1 & 1 & 1 & 1 \\
$A_{2}$ & 1 & 1 & 1 & -1 & -1 \\
$B_{1}$ & 1 & -1 & 1 & 1 & -1 \\
$B_{2}$ & 1 & -1 & 1 & -1 & 1 \\
$E$ & 2 & 0 & -2 & 0 & 0 \\
\hline
\end{tabular} \\ \\

\begin{tabular}{|c | c c c c c c|} 
\hline
\multicolumn{7}{|c|}{$C_{6v}$ Point Group} \\
\hline
$C_{6v}$ & $E$ & $2C_{6}$ & $2C_{3}$ & $C_{2}$ & $3\sigma_{y}$ & $3\sigma_{x}$ \\
\hline
$A_{1}$ & 1 & 1 & 1 & 1 & 1 & 1 \\
$A_{2}$ & 1 & 1 & 1 & 1 & -1 & -1 \\
$B_{1}$ & 1 & -1 & 1 & -1 & 1 & -1 \\
$B_{2}$ & 1 & -1 & 1 & -1 & -1 & 1 \\
$E_{1}$ & 2 & 1 & -1 & -2 & 0 & 0 \\
$E_{2}$ & 2 & -1 & -1 & 2 & 0 & 0 \\
\hline
\end{tabular} \\

The modes discussed in this paper in square and hexagonal nanopatch cavities correspond to the two-dimensional $E$ and $E_{1}$ representations of the $C_{4v}$ and $C_{6v}$ point groups, respectively, and are therefore each doubly degenerate.

\newpage

\listoffigures

\newpage

\begin{IEEEbiography}{Christina Manolatou}
Christina Manolatou was born in Athens, Greece. She received the Ph.D. degree from the Massachusetts Institute of Technology (MIT), Cambridge, in 2001, focusing on theoretical analysis and numerical modeling of integrated optical devices. After a postdoctoral position at MIT, she went to Cornell University, Ithaca, NY, as a Research Scientist. She is the author of the book {\em Passive Components for Dense Optical Integration} published by Kluwer in 2001.   
\end{IEEEbiography}

\begin{IEEEbiography}{Farhan Rana}
Farhan Rana obtained the BS, MS, and PhD degrees all in Electrical Enginnering from the Massachusetts Institute of Technology (MIT), Cambridge, MA (USA). He worked on a variety of different topics related to semiconductor optoelectronics, quantum optics, and mesoscopic physics during his PhD research. Before starting PhD, he worked at IBM's T. J. Watson Research Center on Silicon nanocrystal and quantum dot memory devices. He finished PhD in 2003 and joined the faculty of Electrical and Computer Engineering at Cornell University, NY. His current researh focuses on semiconductor optoelctronics and terahertz devices. He is the recepient of the US National Science Foundation Faculty CAREER award in 2004.      
\end{IEEEbiography}

\end{document}